\documentclass[times]{elsarticle}

\usepackage{jcomp}
\usepackage{framed,multirow}

\usepackage{amssymb}
\usepackage{latexsym}

\usepackage{url}
\usepackage{xcolor}

\usepackage{graphicx}
\usepackage{epstopdf, epsfig}
\usepackage{subfigure}
\usepackage{float}
\usepackage{dcolumn}
\usepackage{bm}
\usepackage{mathrsfs}
\usepackage{mathtools}
\usepackage{lineno}
\usepackage{color}
\usepackage{xcolor}
\usepackage{amsmath,amsfonts,amsbsy,amssymb,array,graphicx,psfrag}

\newcommand{\rone}[1]{{\color{black}{#1}}}

\newcommand{\rthree}[1]{{\color{black}{#1}}}

\definecolor{newcolor}{rgb}{.8,.349,.1}

\begin{document}


\begin{frontmatter}

\title{\rthree{RANS} Turbulence Model Development using CFD-Driven Machine Learning}%

\author[1]{Yaomin Zhao\corref{cor1}}
\cortext[cor1]{Corresponding author:
  yaomin.zhao@unimelb.edu.au}
\author[1]{Harshal D. Akolekar}
\author[1]{Jack Weatheritt}
\author[2]{Vittorio Michelassi}

\author[1]{Richard D. Sandberg}

\address[1]{Department of Mechanical Engineering, University of Melbourne, VIC 3010, Australia}
\address[2]{Baker Hughes, a GE Company, Via Felice Matteucci 2, Firenze 50127, Italy}


\begin{abstract}
  This paper presents a novel CFD-driven machine learning framework to develop Reynolds-averaged Navier-Stokes (RANS) models.
  The CFD-driven training is an extension of the gene expression programming method (Weatheritt \& Sandberg, \textit{J. Comput. Phys.}, \textbf{325}, 22-37 (2016)), but crucially the fitness of candidate models is now evaluated by running RANS calculations in an integrated way, rather than using an algebraic function. Unlike other data-driven methods that fit the Reynolds stresses of trained models to high-fidelity data, the cost function for the CFD-driven training can be defined based on any flow feature from the CFD results. This extends the applicability of the method especially when the training data is limited.
  Furthermore, the resulting model, which is the one providing the most accurate CFD results at the end of the training, inherently shows good performance in RANS calculations.
  To demonstrate the potential of this new method, the CFD-driven machine learning approach is applied to model development for wake mixing in turbomachines.
  A new model is trained based on a high-pressure turbine case and then tested for three additional cases, all representative of modern turbine nozzles.
  \rthree{Despite the geometric configurations and operating conditions being different} among the cases, the predicted wake mixing profiles are significantly improved in all of these \textit{a posteriori} tests.
	Moreover, the model equation is explicitly given and available for analysis, thus it could be deduced that the enhanced wake prediction is predominantly due to the extra diffusion introduced by the CFD-driven model.
\end{abstract}


\end{frontmatter}


\section{Introduction}
Reynolds-averaged Navier-Stokes (RANS) remains the primary tool in most industrial applications due to its low computational cost compared to high-fidelity (Hi-Fi) simulations such as direct numerical simulation (DNS) and large-eddy simulation (LES).
However, RANS suffers from low-fidelity in prediction especially for complex geometries and flow fields~\cite{hunt2005guidelines}, which is mainly due to the limiting assumptions used to model the Reynolds stress tensor $\tau_{ij}$.
Therefore, there has been continuous interest in both academia and industry to improve the predictive accuracy of RANS models routinely applied to aerodynamic and aerothermal design verifications, especially in the aerospace field.

Recently, machine learning has gained popularity in turbulence modelling~\cite{duraisamy2018turbulence}. Various machine-learning techniques have been applied to modify traditional RANS models to enhance the flow prediction, including modifying model parameters with Bayesian methods~\cite{edeling2014bayesian}, introducing a correction factor for the turbulence production term using neural networks~\cite{zhang2015machine}, and adding a spatially distributed correction field via field inversion and Gaussian Process~\cite{parish2016paradigm}, etc.
However, the various types of corrections trained for traditional RANS models are usually not physically interpretable.

Other than simply adding corrections to existing model parameters, more comprehensive efforts have been made to construct new Reynolds stress closures via physics-informed machine learning.
In particular, many studies have focused on improving the  Boussinesq hypothesis
\begin{equation}
\begin{aligned}
\tau_{ij}=\frac{2}{3}\rho k \delta_{ij}-2\mu_tS'_{ij},
\label{eq:boussinesq}
\end{aligned}
\end{equation}
which assumes a linear relation between the anisotropy of the Reynolds stress $a_{ij} = \tau_{ij}-\frac{2}{3}\rho k \delta_{ij}$ and the deviatoric strain-rate tensor $S'_{ij}$.
Such linear relation only holds valid in limited portions of the flow field, while it is questionable in complex flow topologies.
Here, $\rho$ denotes density, $k$ denotes turbulence kinetic energy, and $\mu_t$ represents turbulence viscosity.
A random forest method was used in Wang \emph{et al}.~\cite{wang2017physics} to train Reynolds stress discrepancy functions based on the $a_{ij}$ tensor eigenvalues. The trained discrepancy functions were applied in test cases to show improved prediction of $\tau_{ij}$, but the performance of mean quantities needs to be further validated.
With basis tensors and scalar invariants decomposed from $a_{ij}$, Ling \emph{et al}.~\cite{ling2016reynolds} used a deep neural network to train the Reynolds stress tensor conserving Galilean invariance.
Nevertheless, the resulting neural network from the training is not straightforward to implement into RANS solvers.

In a series of studies, the gene-expression programming (GEP) method has been introduced~\cite{weatheritt2016novel} to develop explicit algebraic (Reynolds) stress models (EASM) based on the stress tensor decomposition proposed by Pope~\cite{pope1975more}.
EASM-like turbulence closures are represented in the form
\begin{equation}
\begin{aligned}
a_{ij} = \sum_{k} f^{(k)}(I_1,I_2, ..., I_n)V_{ij}^k,
\end{aligned}
\label{eq:EASM}
\end{equation}
which can be taken as a solution to the shortcomings of the Boussinesq hypothesis.
This implies the non-dimensionalized anisotropic stress tensor $a_{ij} $ is represented by a combination of tensor bases $V^k_{ij}$ and scalar invariants $I_j$.
Models following Eqn.~(\ref{eq:EASM}) can be explicitly provided through symbolic regression by GEP, which is used to determine the coefficients $f^{(k)}$ based on training data.
Applied in RANS solvers, these GEP-trained models have shown improved predictive accuracy in several \textit{a posteriori} tests such as rectangular ducts~\cite{weatheritt2017machine} and turbomachinery flows~\cite{weatheritt2017development,akolekar2018}.

Although machine learning for turbulence model development is becoming a growing trend, obstacles still exist in training and implementing the models to engineering applications.
One of the major concerns is that most of the training methods aim at accurately modelling the Reynolds stress tensor from Hi-Fi data, while the inherent inconsistency between RANS modelling and high-fidelity data has been neglected~\cite{duraisamy2018turbulence}.
As shown in previous studies~\cite{thompson2016methodology,poroseva2016accuracy}, RANS prediction can remain unsatisfactory even if Reynolds stresses from DNS are used in the CFD calculation.
Such difficulty was also documented by Parneix \emph{et al}.~\cite{parneix1996second,parneix1998procedure}, who tuned a full Reynolds stress model with the aid of the DNS of a backward facing step, but the application of the tuned model to RANS did not confirm the expected improvements.
Therefore, it is questionable whether existing model development methods targeting only the accurate modelling of $\tau_{ij}$ from Hi-Fi data can provide accurate mean flow fields.
Rather than training models exclusively based on high-fidelity data, model development strategies more adaptive to RANS calculations are needed to provide models applicable for pragmatic CFD.

In order to train models adaptive to RANS environments, a novel model development framework named CFD-driven training is introduced in the present work.
By integrating RANS calculations in the GEP model training method introduced by Weatheritt and Sandberg~\cite{weatheritt2016novel}, the objective is to develop Reynolds stress closures of the form given in Eqn.~(\ref{eq:EASM}), which are i) straightforward to be implemented in RANS solvers; ii) robust in a sense of providing stable solutions in RANS calculations; and iii) showing improved predictive accuracy in different \textit{a posteriori} tests.

The present paper is structured as follows. The CFD-driven training frameworks along with the basics for symbolic regression via the GEP method are introduced in Sec.~\ref{sec:method}.
Then the CFD-driven training is applied to turbine wake mixing cases in Sec.~\ref{sec:result}, which includes a brief description of the case setups and training data, the details of the model development process, and the analysis of the predictive results of the trained model.
Furthermore, the trained model is validated for several other cases, and the generalizability of the CFD-dirven model development is discussed in Sec.~\ref{sec:discussion}.
Finally, conclusions are provided in Sec.~\ref{sec:conclusion}.

\section{Methodology}
\label{sec:method}
\subsection{Baseline model}
The model development in the present study is based on the widely applied $k-\omega$ SST model~\cite{menter1994two}, in which the equations for the turbulent kinetic energy $k$ and the specific dissipation rate $\omega$ are written as
\begin{equation}
  \begin{aligned}
    \frac{\mathrm{D} \rho k}{\mathrm{D} t}&=\tau_{i j} \frac{\partial u_{i}}{\partial x_{j}}-\beta^{*} \rho \omega k+\frac{\partial}{\partial x_{j}}\left[\left(\mu+\sigma_{k} \mu_{t}\right) \frac{\partial k}{\partial x_{j}}\right],\\
    \frac{\mathrm{D} \rho \omega}{\mathrm{D} t}&=\frac{\alpha}{\nu_{t}} \tau_{i j} \frac{\partial u_{i}}{\partial x_{j}}-\beta \rho \omega^{2}+\frac{\partial}{\partial x_{j}}\left[\left(\mu+\sigma_{\omega} \mu_{t}\right) \frac{\partial \omega}{\partial x_{j}}\right]+2\left(1-F_{1}\right) \rho \sigma_{\omega 2} \frac{1}{\omega} \frac{\partial k}{\partial x_{j}} \frac{\partial \omega}{\partial x_{j}},
  \end{aligned}
  \label{eq:SST}
\end{equation}
where $u_i$ is the velocity component.
Other relations to close the transport equations are
\begin{equation}
  \begin{aligned}
    &F_{1}=\tanh \left[\left(\min \left[\max \left(\frac{\sqrt{k}}{\beta^{*} \omega y}, \frac{500 v}{y^{2} \omega}\right), \frac{4 \sigma_{\omega 2} k}{C D_{k \omega} y^{2}}\right]\right)^{4}\right],\\
    &F_{2}=\tanh \left[\left[\max \left(\frac{2 \sqrt{k}}{\beta^{*} \omega y}, \frac{500 v}{y^{2} \omega}\right)\right]^{2}\right],
    C D_{k \omega}=\max \left(2 \sigma_{\omega 2} \frac{1}{\omega} \frac{\partial k}{\partial x_{i}} \frac{\partial \omega}{\partial x_{i}}, 10^{-10}\right).
  \end{aligned}
\end{equation}
The eddy viscosity is given as
\begin{equation}
\nu_{t}=\frac{a_{1} k}{\max \left(a_{1} \omega, \Omega F_{2}\right)}.
\end{equation}

Furthermore, the model coefficients (say $\psi$) are obtained by blending the coefficients of the $k-\epsilon$ and $k-\omega$ models as $\psi=\psi_{2}+F_{1}\left(\psi_{1}-\psi_{2}\right)$, and the whole set of the coefficients are
\begin{equation}
  \begin{aligned}
&\alpha_{1}=0.553,\quad \alpha_{2}=0.44, \quad \beta_{1}=0.075, \quad \beta_{2}=0.0828 \\ 
&\sigma_{k1}=0.85, \quad \sigma_{k2}=1.0, \quad\sigma_{\omega 1}=0.5, \quad \sigma_{\omega 2}=0.856,   \\ 
&\beta^{*}=0.09, \quad \kappa=0.41, \quad a_1=0.31.
\end{aligned}
\end{equation}

\rone{To simulate the laminar-turbulent transition in the boundary layer, the well-established $\gamma$-$Re_{\theta}$ transition model \cite{langtry2009correlation} is used in the RANS calculations. 
The transport equations for the intermittency factor $\gamma$ and the transition momentum thickness Reynolds number $\tilde{R} e_{\theta t}$ read:
\begin{equation}
  \begin{aligned}
&\frac{\partial(\rho \gamma)}{\partial t}+\frac{\partial\left(\rho U_{j} \gamma\right)}{\partial x_{j}}=F_{\text {length }} c_{a 1} \rho S\left[\gamma F_{\text {onset }}\right]^{0.5}\left(1-c_{e 1} \gamma\right)-c_{a 2} \rho \Omega \gamma F_{\text {turb }}\left(c_{e 2} \gamma-1\right)+\frac{\partial}{\partial x_{j}}\left[\left(\mu+\frac{\mu_{t}}{\sigma_{f}}\right) \frac{\partial \gamma}{\partial x_{j}}\right],\\
&\frac{\partial\left(\rho \tilde{R} e_{\theta t}\right)}{\partial t}+\frac{\partial\left(\rho U_{j} \tilde{R} e_{\theta t}\right)}{\partial x_{j}}=c_{\theta t} \frac{\rho^2 U^{2}}{500 \mu} \left(R e_{\theta t}-\tilde{R} e_{\theta t}\right)\left(1.0-F_{\theta t}\right)+\frac{\partial}{\partial x_{j}}\left[\sigma_{\theta t}\left(\mu+\mu_{t}\right) \frac{\partial \tilde{R} e_{\theta t}}{\partial x_{j}}\right].
  \end{aligned}
\end{equation}
A brief list of the parameters related to the transition models is given below. We remark that for details on the selection of these parameters used in this correlation-based model the reader is referred to Langtry and Menter \cite{langtry2009correlation}.
\begin{equation}
  \begin{aligned}
&R e_{V}=\frac{\rho y^{2} S}{\mu},\quad F_{\text {onsetl }}=\frac{R e_{v}}{2.193 \cdot R e_{\theta c}}, \quad F_{\text {onset } 2}=\min \left(\max \left(F_{\text {onset } 1}, F_{\text {onset }}^{4}\right), 2.0\right) \\ 
&R_{T}=\frac{\rho k}{\mu \omega}, \quad F_{\text {onset } 3}=\max \left(1-\left(\frac{R_{T}}{2.5}\right)^{3}, 0\right), \quad F_{\text {onset }}=\max \left(F_{\text {onset } 2}-F_{\text {onset } 3}, 0\right),\\
&c_{e 1}=1.0, \quad c_{a 1}=2,\quad c_{e 2}=50, \quad c_{a 2}=0.06, \quad \sigma_{f}=1.0,\\
&F_{\theta t}=\min \left(\max \left(F_{\text {wake }} \cdot e^{-\left(\frac{y}{\delta}\right)^{4}}, 1.0-\left(\frac{\gamma-1 / c_{e 2}}{1.0-1 / c_{e 2}}\right)^{2}\right), 1.0\right)\\
&R e_{\omega}=\frac{\rho \omega y^{2}}{\mu}, \quad F_{\text {wake }}=e^{-\left(\frac{R e \omega}{1 E+\xi}\right)^{2}},\quad c_{\theta t}=0.03, \quad \sigma_{\theta t}=2.0.
\end{aligned}
\end{equation}

Furthermore, the transition model interacts with the orginal SST model in Eqn.~(\ref{eq:SST}) as:
\begin{equation}
\begin{aligned}
&\frac{\partial}{\partial t}(\rho k)+\frac{\partial}{\partial x_{j}}\left(\rho u_{j} k\right)=\tilde{P}_{k}-\tilde{D}_{k}+\frac{\partial}{\partial x_{j}}\left(\left(\mu+\sigma_{k} \mu_{t}\right) \frac{\partial k}{\partial x_{j}}\right);\\
&\tilde{P}_{k}=\gamma_{\mathrm{eff}} P_{k}, \quad \tilde{D}_{k}=\min \left(\max \left(\gamma_{\mathrm{eff}}, 0.1\right), 1.0\right) D_{k},\\
&R_{y}=\frac{\rho y \sqrt{k}}{\mu}, \quad F_{3}=e^{-\left(\frac{R_{y}}{120}\right)^{8}}, \quad F_{1}=\max \left(F_{\text {lorig }}, F_{3}\right).
\end{aligned}
\end{equation}
Note that $P_{k}=\tau_{i j} \frac{\partial u_{i}}{\partial x_{j}}$ and $D_{k}=\beta^{*} \rho \omega k$ are the original production and dissipation terms in Eqn.~(\ref{eq:SST}).
Also, it is worth pointing out that the transition model is switched off outside the boundary layer, especially in the wake region which the model training in the present study focuses on.
}

We want to stress here that the baseline model applied in the present study is based on the Boussinesq approximation as in Eqn.~(\ref{eq:boussinesq}), which assumes a linear relation between the anisotropy of the Reynolds stress and the deviatoric strain-rate tensor.
In the current study, extra anisotropic terms will be added via data-driven methods, to \rthree{improve} the predictive accuracy as shown in the following section. 

\subsection{GEP for model development}
GEP, first introduced for turbulence model development by Weatheritt and Sandberg~\cite{weatheritt2016novel}, is applied in the present study to train an EASM-like Reynolds stress relation by adding an extra anisotropic stress $a_{ij}^{x}$ as
\begin{equation}
\begin{aligned}
\tau_{ij}= \frac{2}{3}\rho k \delta_{ij} - 2\mu_tS'_{ij}+a_{ij}^{x}.
\end{aligned}
\end{equation}
The $a_{ij}^{x}$ is represented as a linear combination of tensor bases $V^k_{ij}$ and scalar invariants $I_k$ as in Eqn.~(\ref{eq:EASM}).
To focus on statistically two-dimensional flows which are representative of the mid-span section of turbine blades, the invariants and basis tensors considered in the present work are
\begin{equation}
\begin{aligned}
&V_{ij}^1 = s_{ij}, V^2_{ij}=s_{ik}\omega_{kj}-\omega_{ik}s_{kj},\\
&V^3_{ij} =s_{ik}s_{kj}-\frac{1}{3}\delta_{ij}s_{mn}s_{nm},\\
&I_1 =s_{mn}s_{nm}, I_2=\omega_{mn}\omega_{nm}.
\end{aligned}
\end{equation}
Here, $V^k_{ij}$ and $I_k$ are functions of the non-dimensionalized strain rate tensor $s_{ij}=t_IS'_{ij}$ and the rotation rate tensor $\omega_{ij}=t_I\Omega_{ij}$.
The turbulence time scale is computed as  $t_I=\frac{1}{\omega}$, with $\omega$ denoting the specific dissipation rate.
It is noted that while the dissipation rate can be resolved in high-fidelity simulations, a separate scalar equation, e.g. the $\omega$ \rthree{equation}, is usually needed to provide the turbulence scales for RANS calculations.

The GEP model training starts from an initial population consisting of many candidate EASMs $a_{ij}^{GEP}$, \rthree{in which the tensor bases $V^k_{ij}$ are fixed while the functional forms for the coefficents $f^{(k)}$ in Eqn.~(\ref{eq:EASM}) depend on $I_k$ and need to be explored}. These inital model functions are randomly generated and therefore have distinct functional forms and random constants. \rthree{Each individual candidate model consist of three genes that represent separate coefficients for $V^k_{ij}$, and the length of the expression of the coefficient function is dominated by the head length in the gene and the truncation size of the expression tree. In the present study, the head length is set as $5$ while the maximum size of the expression tree is set as $23$. For more details about the setting of the GEP parameters, the reader is referred to Weatheritt and Sandberg~\cite{weatheritt2016novel}.}
As presented in Weatheritt and Sandberg~\cite{weatheritt2016novel}, the final models derived from 100 different training runs with varying initial populations share syntactically similar functions. Therefore, the model training with GEP is not significantly dependent on the initial condition, as long as the population size is sufficiently large.
In the present study, the population size is set as $1000$, and three training runs with different initial populations eventually converge to functionally similar expressions.

The population of models evolves in the GEP iterations by genetic operations such as mutations and combinations until the best model in the generation provides an acceptable approximation to pre-set target functions.
Through this non-deterministic evolution by GEP, symbolic regression is achieved to find an optimal model that best fits the training data~\cite{ferreira2001algorithm}.
While more details about the GEP method are given in \cite{weatheritt2016novel}, we want to stress here that the model development characteristics are mainly determined by the selection process.
The fitness of every generated model is evaluated by its cost function $J(a_{ij}^{GEP})$, which describes the deviation between the model and training data.
The general principal for the definition of the cost function is that candidate models that better fit training data should have lower cost values, which result in better chances to be selected for the next generation.
We note that the specific form of the cost function usually depends on the training framework and the cases, which will be discussed in detail in Sec.~\ref{sec:framework} and Sec.~\ref{sec:cases}.

\subsection{Training frameworks}
\label{sec:framework}
The training framework, in particular how to evaluate candidate models during evolution, is critical to the quality of the final model.
Previous GEP studies~\cite{weatheritt2016novel,weatheritt2017development,akolekar2018}, along with other existing machine-learning driven model development work~\cite[e.g.][]{ling2016reynolds,parish2016paradigm}, typically evaluate the models depending on time-averaged Hi-Fi data.
This is denoted as `frozen' training in the present study, with the `frozen' here meaning the pre-processed Hi-Fi data remains unchanged in the training.
It is noted that the inherent inconsistency between Hi-Fi data and RANS environment, such as discrepancies of the turbulent dissipation rate providing turbulence scales for non-dimensionalization, limits the applicability of the models trained with frozen training~\cite{duraisamy2018turbulence}.
Therefore, a novel framework that includes CFD calculations in model training iterations is proposed in the present study.
The CFD-driven training, aiming to train models well adapted to the RANS mean velocity and turbulence scales, is introduced following a brief description of the frozen training in this section.

\subsubsection{Frozen training}
For frozen training, candidate models generated by GEP are evaluated based on Hi-Fi data.
With the tensor bases $V_{ij}^k$ and scalar invariants $I_k$ processed from a Hi-Fi flow field, the $a_{ij}^{GEP}$ can be calculated using the candidate model equations from GEP.
On the other hand, the $a_{ij}^{HiFi}$ directly obtained from Hi-Fi data is considered as `accurate' Reynolds stress anisotropy.
The cost functions $J^{fro}$ are defined by the deviation between $a_{ij}^{GEP}$ and $a_{ij}^{HiFi}$.
An example used in the present study is
\begin{equation}
J^{fro} = \frac{1}{N}\sum_{n=1}^N\sum_{i,j}\frac{|a_{ij}^{HiFi}-a_{ij}^{GEP}|}{|a_{ij}^{HiFi}|},
\label{eq:fitness}
\end{equation}
where $N$ denotes the number of training data points.
We can infer from Eqn.~(\ref{eq:fitness}) that candidate models which better fit Hi-Fi data will have a lower cost.
The cost functions are then used to direct the selection and reproduction of the models.
Through a series of iterations, the final model with minimized cost functions will be obtained.

We can see that the models are trained exclusively based on Hi-Fi data.
The frozen-trained model generally provides accurate $a_{ij}$ predictions as long as Hi-Fi basis tensors and scalar invariants are available as input.
However, once the model is implemented into RANS, good performance is not guaranteed. In fact RANS will predict a flow field that may deviate from the Hi-Fi flow field. The impact of such variations on the GEP trained model is non-linear and difficult to forecast.

\subsubsection{CFD-driven training}
\begin{figure}
\begin{center}
  \includegraphics[width=1.0\textwidth]{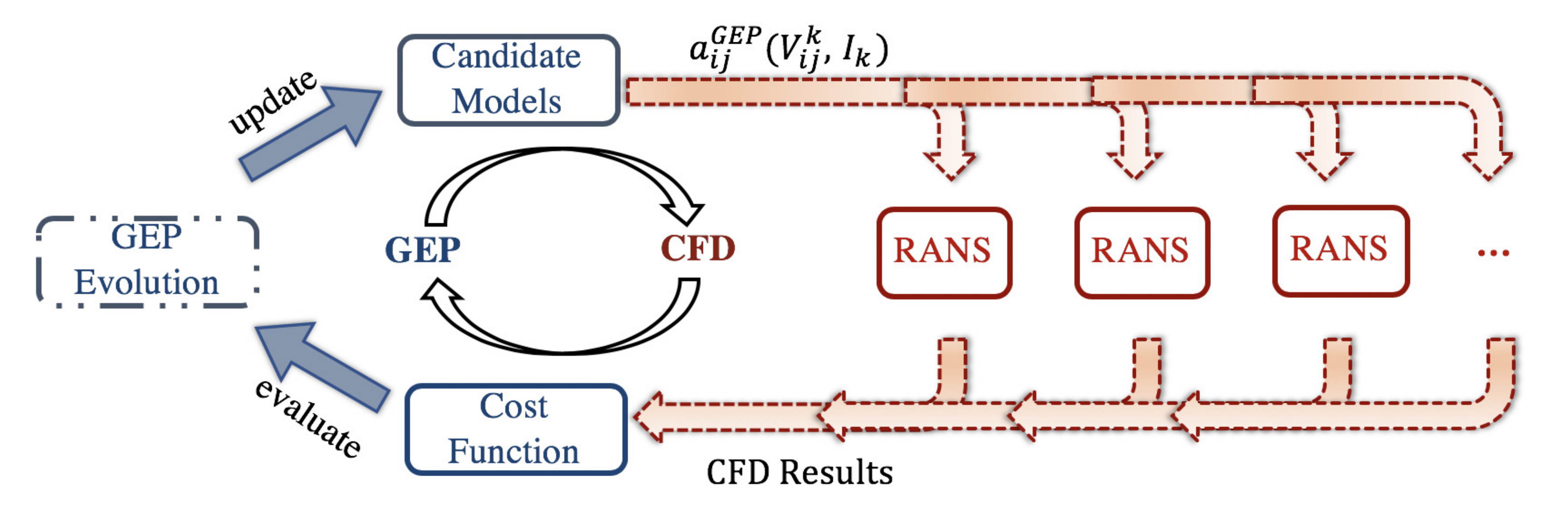}
  \caption{Schematic for CFD-driven training. The CFD calculations represented by red symbols are integrated in the GEP training loop, taking turbulence stress model equations as input and providing cost functions of candidate models for the evolution algorithm.}
\label{fig:CFD_in_loop}
\end{center}
\end{figure}

As shown by the schematic for the CFD-driven training in Fig.~\ref{fig:CFD_in_loop}, the candidate models are evaluated by running RANS calculations that will provide feedback in the GEP training loop.
For every generation of candidate models, the model equations are read into the RANS solver, and then a series of real-time CFD calculations are performed to test different models in parallel.
\rone{This process is fully automatic, and the procedure of each of these CFD calculations can be summarized as follows:
\begin{enumerate}
  \item The string of the candidate model equation is written to a file, and this file is then read by the RANS solver;
  \item The RANS solver takes the GEP equation as the turbulence model and begins iteration;
  \item The CFD iteration ends if 
  \begin{itemize}
     \item convergence is reached: cost function can be evaluated based on the comparison between the RANS results and the Hi-Fi data;
     \item or the model is not converging: cost function is set to be infinity or an extremely large value.
   \end{itemize}  
\end{enumerate}}
Therefore, the cost functions of the candidate models can be quantified after each RANS calculation is finished, and the Hi-Fi data is only used to be compared to the converged RANS results.
Based on the cost functions of the model functions, a new generation of models is formed through GEP evolution, and then the updated models need to be evaluated again through the integrated CFD calculations. This training process iterates and the population of models evolves, until the best model in the generation provides an acceptable cost value.

We remark that \rthree{although the idea of integrating CFD calculations in an optimization process is well accepted in the context of uncertainty quantification analysis applied to RANS modeling \cite{xiao2019}, the present study is novel in that it combines the CFD calculations with machine learning methods to train RANS models, rather than calibrating existing models.}
The CFD-driven training is executable with the GEP method because the model equations are explicitly given and can be instantly implemented into RANS solvers during training.
With RANS calculations performed in the training loop, the resulting CFD-driven models thus are also ready to be implemented into industrial design tools.
Another advantage of the CFD-driven training is that the definition of the cost function is more flexible compared to the frozen training.
Rather than being restricted to quantities that are part of the closure terms in frozen training, the CFD-driven cost functions can be defined based on any important flow feature of interest to designers.
As an example, the case setup and the cost function for enhanced wake mixing predictions will be shown in Sec.~\ref{sec:model}.

\section{Numerical Setup and Results}
\label{sec:result}
\subsection{Case setup for turbine wake mixing}
\label{sec:cases}
As key components of turbomachines, high-pressure turbine (HPT) and low-pressure turbine (LPT) flows are challenging for RANS calculations due to their extremely complex flow features.
In particular, wakes in turbine nozzles are important due to their contribution to unsteady losses, both in the boundary layer and the freestream flow, for aeromechanics, and because of their possible perturbations to film cooling in HPT. An incorrect prediction of the wake decay may be responsible for important design weaknesses~\cite{pichler2016investigation}.

\begin{figure}
\begin{center}
  \includegraphics[width=0.9\textwidth]{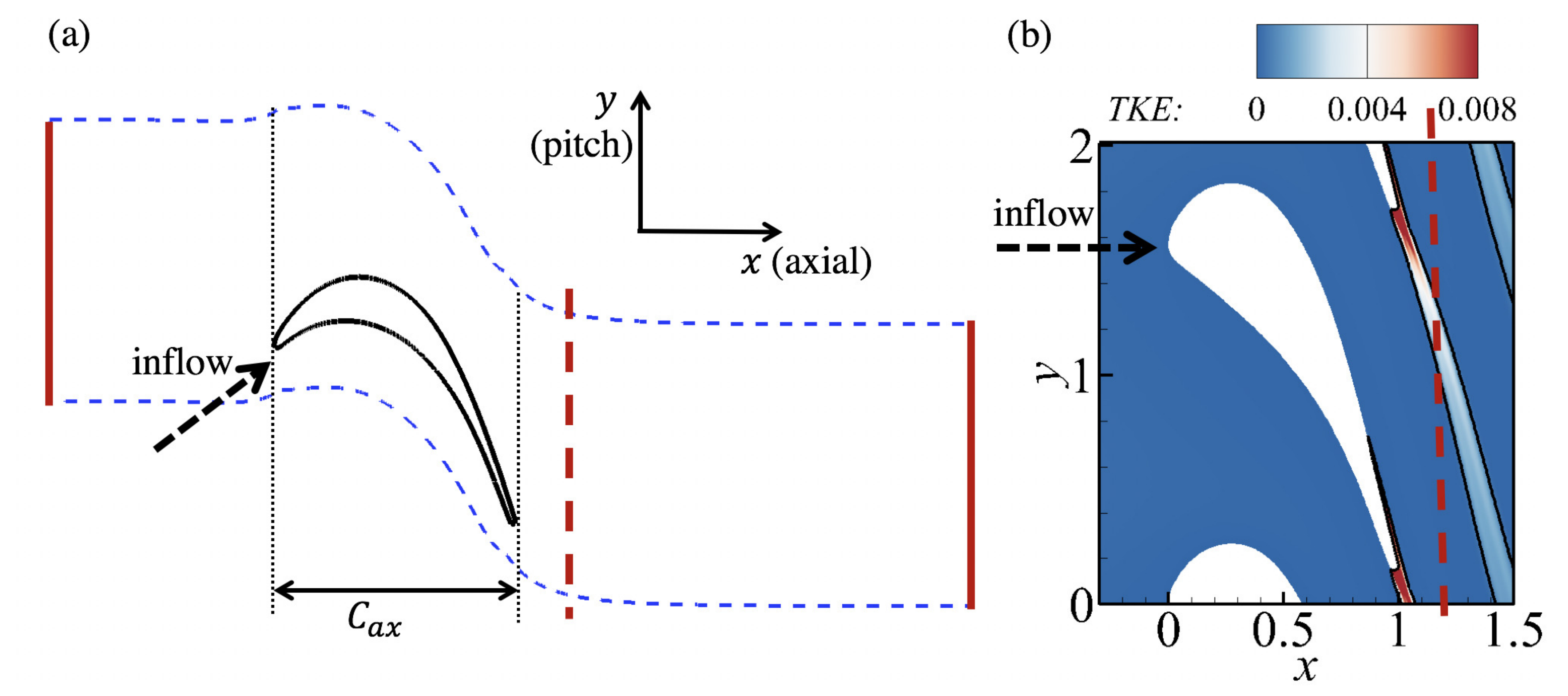}
  \caption{Setup for cases: (a) LPT blade geometry; (b) HPT blade geometry and the wake region indicated by contour of turbulent kinetic energy, with the blade leading edge at $x=0$.}
\label{fig:cases}
\end{center}
\end{figure}

Four different turbine cases which are representative of low and high pressure turbine nozzles for modern aircraft engines, including the VKI LS89 HPT~\cite{arts1990aero} and the T106A LPT~\cite{stadtmuller2001test} at different Reynolds numbers, are selected to test the CFD-driven training framework.
The geometries of the HPT and LPT blades are shown in Fig.~\ref{fig:cases}, with periodic boundary conditions implemented in the pitchwise direction denoted by $y$, and inflow and outflow conditions applied at the boundaries in the axial direction by $x$.
The blade geometries and the inflow angles indicated by the black dashed arrows in Fig.~\ref{fig:cases} show the considerable differences between the cases.
Furthermore, as listed in table \ref{tab:cases}, the flows in these cases are at different Reynolds and Mach numbers, and the flow features of the boundary layers on the blade surface also deviate significantly.
We use the $Re=0.57\times10^6$ HPT case, denoted as case A, as the training case, while the other three cases are used as testing cases for cross-validation, to assess whether the model trained in case A is suitable for a broad parameter space.

\begin{table}
  \begin{center}
\def~{\hphantom{0}}
  \begin{tabular}{lcccccc}
 	\hline
      $Cases$ &  $Re$             & $Ma_{exit}$ & Flow features      &  O-H Mesh & $\Delta n^+_{wall}$\\[3pt]
    \hline
       HPT A &   $0.57\times10^6$ & ~~0.9    & transition \& shocks  &  $705\times65+101\times161$ & $<2.0$\\
       HPT B  & $1.1\times10^6$  & ~~0.9    & transition \& shocks  &  $705\times65+101\times161$ & $<2.0$\\
       LPT C  & $0.6\times10^5$   & ~~0.4    & transition \& open separation & $711\times101+81\times141$ & $<1.0$\\
       LPT D  & $1.0\times10^5$  & ~~0.4   & transition \& closed separation & $711\times101+81\times141$ & $<1.0$\\
    \hline
  \end{tabular}
  \caption{Parameters for turbine flow test cases.}
  \label{tab:cases}
  \end{center}
\end{table}

The Hi-Fi data for these cases is provided from previous DNS and high-resolution LES using code HiPSTAR~\cite{sandberg2015compressible,michelassi2015compressible,pichler2017high}, and RANS calculations are conducted with the RANS solver TRAF~\cite{arnone1995multigrid}.
In the RANS calculations required for model development and testing, the $k-\omega$ SST with $\gamma-Re_\theta$ transitional correlation~\cite{langtry2009correlation} is applied as the baseline model.
The meshes used in the RANS calculations include an O-type grid around the blade and an H-type grid downstream for the wake region.
A thorough grid independence study has been conducted for the RANS calculations to ensure results are grid-independent~\cite{zhao2019using}, and \rone{this information is provided in appendix A.
The mesh details used in the training and testing cases} are presented in table \ref{tab:cases}.
It is shown by the distance of the first grid point from the wall $\Delta n^+_{wall}$ that the LPT meshes are further refined to accurately predict the separation on the blade.

\subsection{Model development by CFD-driven training}
\label{sec:model}
As traditional RANS models applied in turbine flows share a common weakness in correctly capturing wake mixing~\cite{pichler2016investigation}, the wake profile prediction, which is of critical interest for aerodynamic performance and characteristics of downstream blade boundary layers, is set as the optimisation target.
Accordingly, the GEP cost function for the CFD-driven training is defined as
\begin{equation}
\begin{aligned}
\omega(y)&=\frac{p_{t}^{i}-p_t(y)}{p_{t}^{i}-p^{o}},\\
\Delta_x &=\frac{1}{L_y}\int_0^{L_y}(\frac{\omega_{HiFi}(y)-\omega_{RANS}(y)}{max_y(\omega_{HiFi})})^2dy,\\
J^{CFD} &= \Delta_{x_1}+\Delta_{x_2}.
\end{aligned}
\end{equation}
Here, $\omega(y)$ represents the kinetic loss profile along the pitchwise axis $y$ as represented by the red dashed lines downstream of the blade in Fig.~\ref{fig:cases}, with $p_{t}^{i}$, $p_t^{o}$, and $p^{o}$ denoting inlet total pressure, outlet total pressure, and outlet static pressure, respectively.
Moreover, $\Delta_x$ stands for normalised deviation between the wake loss profiles from Hi-Fi and RANS results.
In order to train models for the whole wake region rather than just a profile at one $x$ location, the final cost function is defined as a summation of  $\Delta_x$ at two axial locations $x_1=1.15C_{ax}$ and $x_2=1.25C_{ax}$, as these two axial locations represent the range of most probable downstream blade leading edge position.
Note that the Hi-Fi data needed for the computation of the CFD-driven cost function $J^{CFD}$ only needs to include the kinetic loss profiles $\omega_{HiFi}(y)$ at these two locations.

To focus on improving model accuracy in wake mixing, training data is extracted from the turbine wake.
As indicated by the turbulence kinetic energy (TKE) contour in Fig.~\ref{fig:cases}(b), the wake region is masked by the criterion $k>5\%k_{max}$.
Moreover, another constraint $x>1.05$ is implemented to avoid the boundary layer and near trailing edge interference~\cite{weatheritt2017development,akolekar2018}.
Note that the trained Reynolds stress closure is only implemented in the masked wake region for both model development and testing, while the $k-\omega$ SST model with $\gamma-Re_\theta$ transitional correlation~\cite{langtry2009correlation} is applied in the rest of the flow domain.

\begin{figure}
  \begin{center}
    \includegraphics[width=0.5\textwidth]{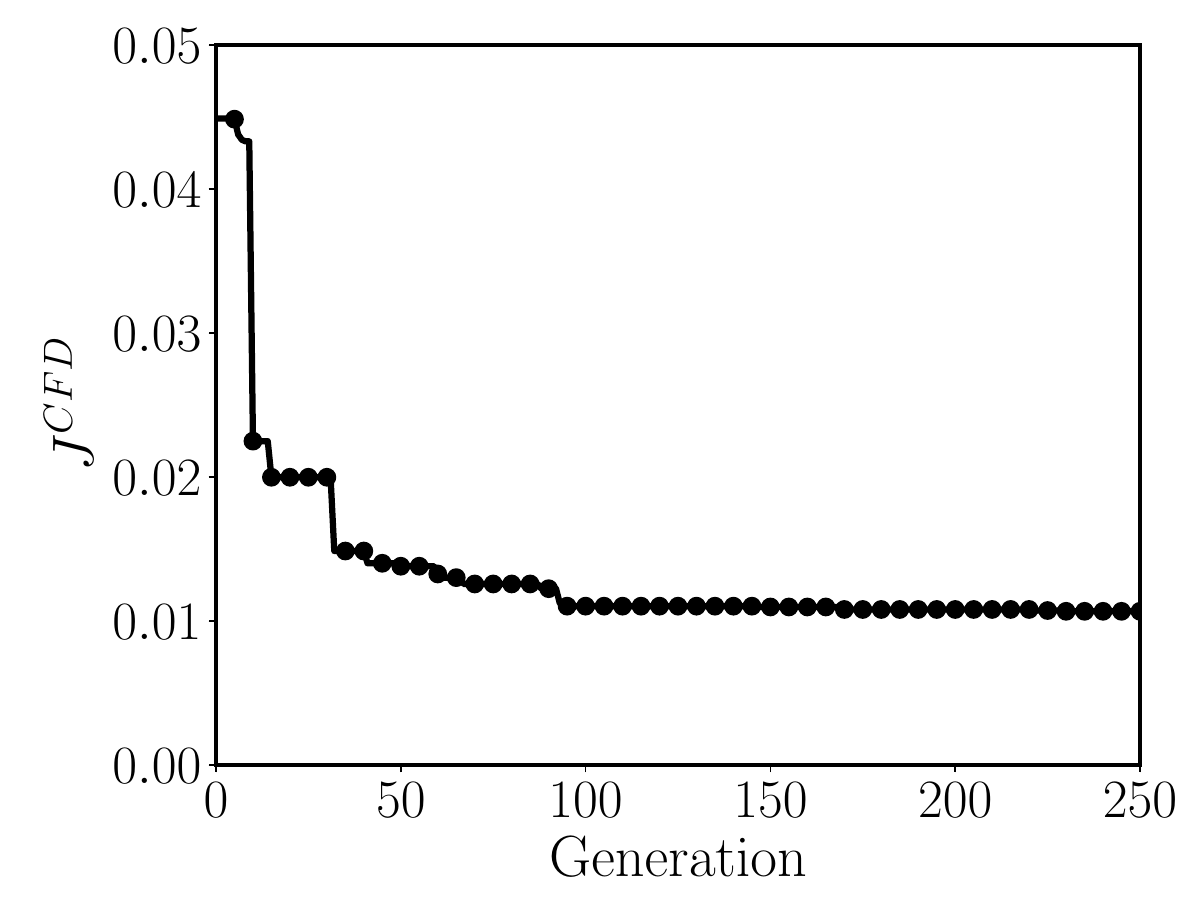}
    \caption{\rthree{Example} of the cost function evolution in training.}
  \label{fig:model_dev}
  \end{center}
\end{figure}

The CFD-driven training framework is applied in the HPT case A to develop a new Reynolds stress closure for turbine wake mixing.
In order to reduce the steps to get a converged solution for the models in the training loop, the CFD calculations required to evaluate the candidate models are initialized from a flow field obtained from the baseline RANS model. In the present study, each CFD calculation runs for 200 steps, and the residual for subiterations are monitored to make sure that the solution is converged with a root-mean-square residual \rone{of the continuity equation} below $10^{-7}$.
For each generation, the cost functions of the candidate models are thus evaluated based on the converged flow fields from the corresponding CFD calculations, according to the definition presented in Eqn.~(\ref{eq:fitness}). Then the model equations evolve to the next generation through genetic operations, until the model is converged or a satisfactory cost function level is achieved. The evolution of $J^{CFD}$ from the best model in the population is presented in Fig.\ref{fig:model_dev}, showing that \rthree{the training gradually converges after approximately one hundred generations.}
Due to the RANS calculations required in the training loop, the CFD-driven training in the present study typically requires $O(10^3)$ CPU hours which is more computationally expensive than the frozen training.
Nevertheless, this is still acceptable as it is only associated with the model development process.
\rone{As the trained model only introduces algebraic extensions to existing equations and no extra equations are solved, the overheads caused by applying the trained model functions in RANS calculations, compared to the baseline linear-model calculations, are around $5\%$ for the present cases.}

The resulting model from the CFD-driven training is presented as $\tau_{ij}^{CFD}$.
In addition, a frozen-trained model $\tau_{ij}^{fro}$, \rthree{which was trained in previous work~\cite{weatheritt2017machine} based on the same high-fidelity data of HPT case A}, is shown for comparison
\begin{equation}
\begin{aligned}
\tau_{ij}^{CFD} &= \frac{2}{3}\rho k \delta_{ij} - 2\mu_tS'_{ij}+2\rho k[\\
&(-2.57+I_1)V_{ij}^1+4.0V_{ij}^2 +(-0.11+0.09I_1I_2+I_1I_2^2 )V_{ij}^3],\\
\tau_{ij}^{fro} &= \frac{2}{3}\rho k \delta_{ij} - 2\mu_tS'_{ij}+2\rho k[\\
&(-1.334 + 0.438I_1 + 2.653I_2+ 0.0102I_1^2-1.021I_2^2+12.280I_1I_2)V_{ij}^1  \\
         & +(0.573 - 1.096I_1 + 8.985I_2 - 0.1102I_1^2 + 2.876I_2^2 +90.633I_1I_2)V_{ij}^2 \\
         & +(12.861 - 25.094I_1 + 6.449I_2 + 1.020I_1^2 - 304.979I_1I_2 - 184.519I_2^2)V_{ij}^3].
\label{eq:model_HPT}
\end{aligned}
\end{equation}
It is noted that the CFD-driven model exhibits a much simpler form compared to the frozen model as the coefficient functions contain less high-order invariants, \rthree{even though the training of the frozen model is also based on the data extracted from the wake region, as shown in Fig.~\ref{fig:cases}(b).}
This is because the higher-order terms in the frozen model, like $90.633I_1I_2$, usually result in very high-amplitude values on some points in RANS calculations, reducing numerical stability.
Therefore, such terms fail to survive in the CFD-driven evolution, and the training via running RANS calculations in the loop tends to produce a more robust model.
As non-linear turbulence model closures are known to come with stability issues, the capability of the CFD driven training to filter out numerically stiff models is very important.

\subsection{Predictive results}
\label{sec:HPT_predict}
\begin{figure}
\centering \subfigure{
  \begin{minipage}[c]{0.48\textwidth}{}
  \includegraphics[width=2.8in]{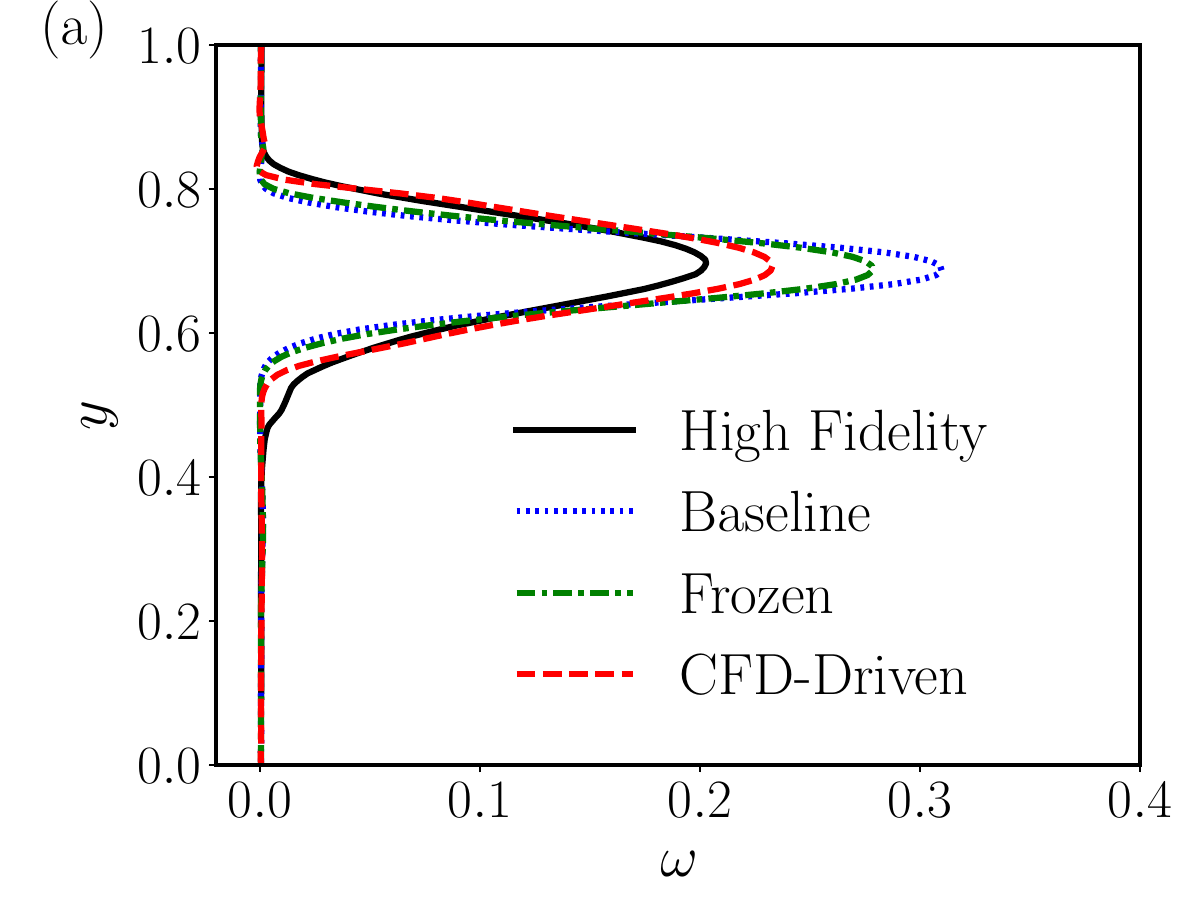}
  \end{minipage}}
\centering \subfigure{
  \begin{minipage}[c]{0.48\textwidth}{}
  \includegraphics[width=2.8in]{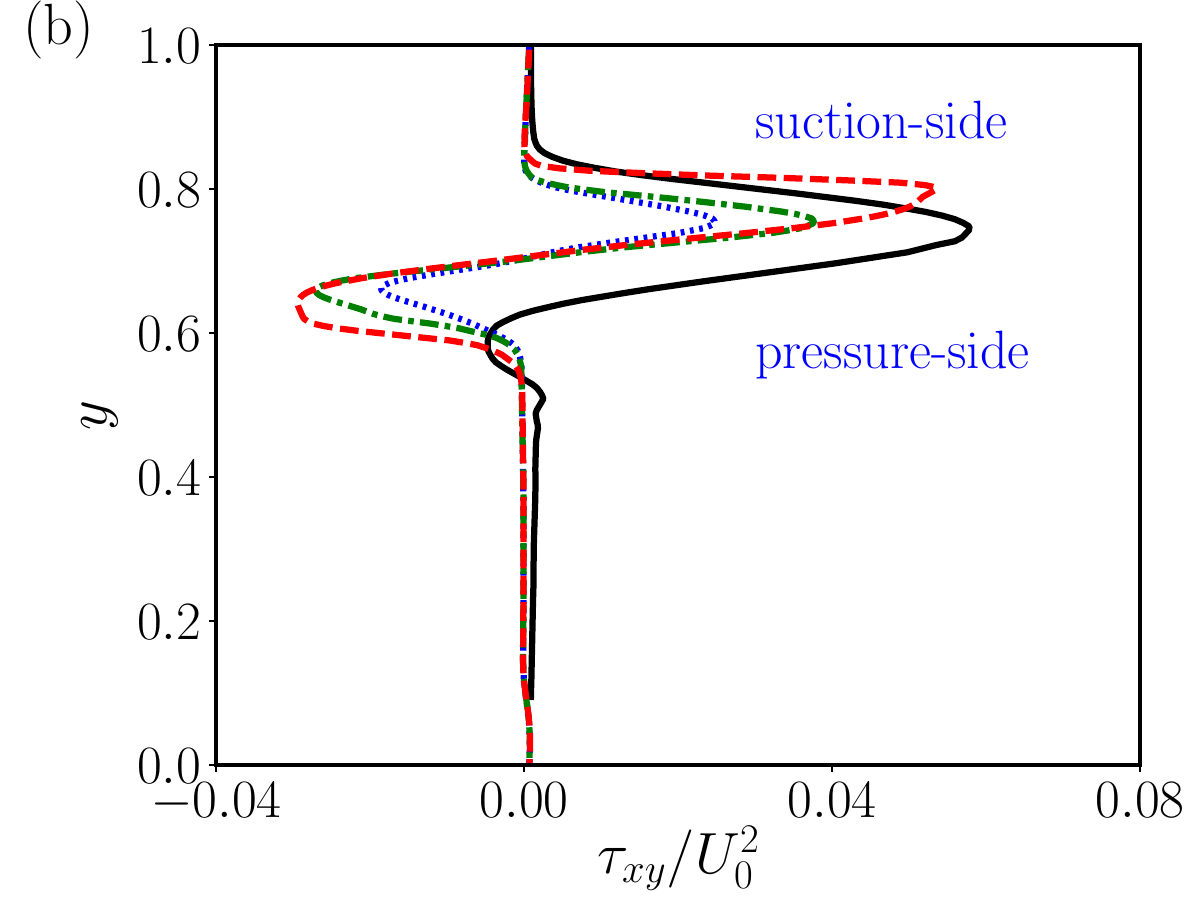}
  \end{minipage}}
  \caption{Profiles at $20\%$ axial chord downstream of blade trailing edge for HPT case A, (a) kinetic loss; (b) shear stress term $\tau_{xy}$.}
\label{fig:HPT_wake}
\end{figure}

The GEP models in Eqn.~(\ref{eq:model_HPT}) are first tested \textit{a posteriori} for the training HPT case A.
The trained model equations are implemented into the RANS solver, and the wake mixing profiles from RANS calculations with these models are shown in Fig.~\ref{fig:HPT_wake}(a).
In addition, profiles from Hi-Fi data and baseline models without any additional stress tensors are presented for comparison.
We can see that the baseline model relying on a linear stress-strain relationship significantly over-predicts the wake peak and under-predicts the wake width.
While the frozen model reduces the wake peak, the CFD-driven model provides an even better agreement with the reference Hi-Fi data, in terms of both the peak value and the profile width.

The wake prediction improvement from the trained models can be analysed by considering the leading-order terms in Eqn.~(\ref{eq:model_HPT}) because the amplitudes of the scalar invariants $I_1$ and $I_2$ are smaller than the constants~\cite{akolekar2018}.
With $\mu_tS'_{ij}=\rho kV_{ij}^1$, we have
\begin{equation}
\begin{aligned}
\tau_{ij}^{CFD-L} &= \frac{2}{3}\rho k \delta_{ij} - 2\mu_t(1.0+ 2.57)S'_{ij},\\
\tau_{ij}^{fro-L} &= \frac{2}{3}\rho k \delta_{ij} - 2\mu_t(1.0+ 1.334)S'_{ij}.
\label{eq:model_simp}
\end{aligned}
\end{equation}
Compared to the baseline model, the frozen model increases diffusion by roughly $133\%$, while the increase of the CFD-driven training is even larger at $257\%$.
With the additional diffusion and extra anisotropic terms from $V_{ij}^2$ and $V_{ij}^3$, the CFD-driven model allows for a stronger mixing and thus a better prediction of the wake profile spreading as seen in Fig.~\ref{fig:HPT_wake}(a).

\begin{figure}
  \centering \subfigure{
    \begin{minipage}[c]{0.48\textwidth}{}
    \includegraphics[width=2.8in]{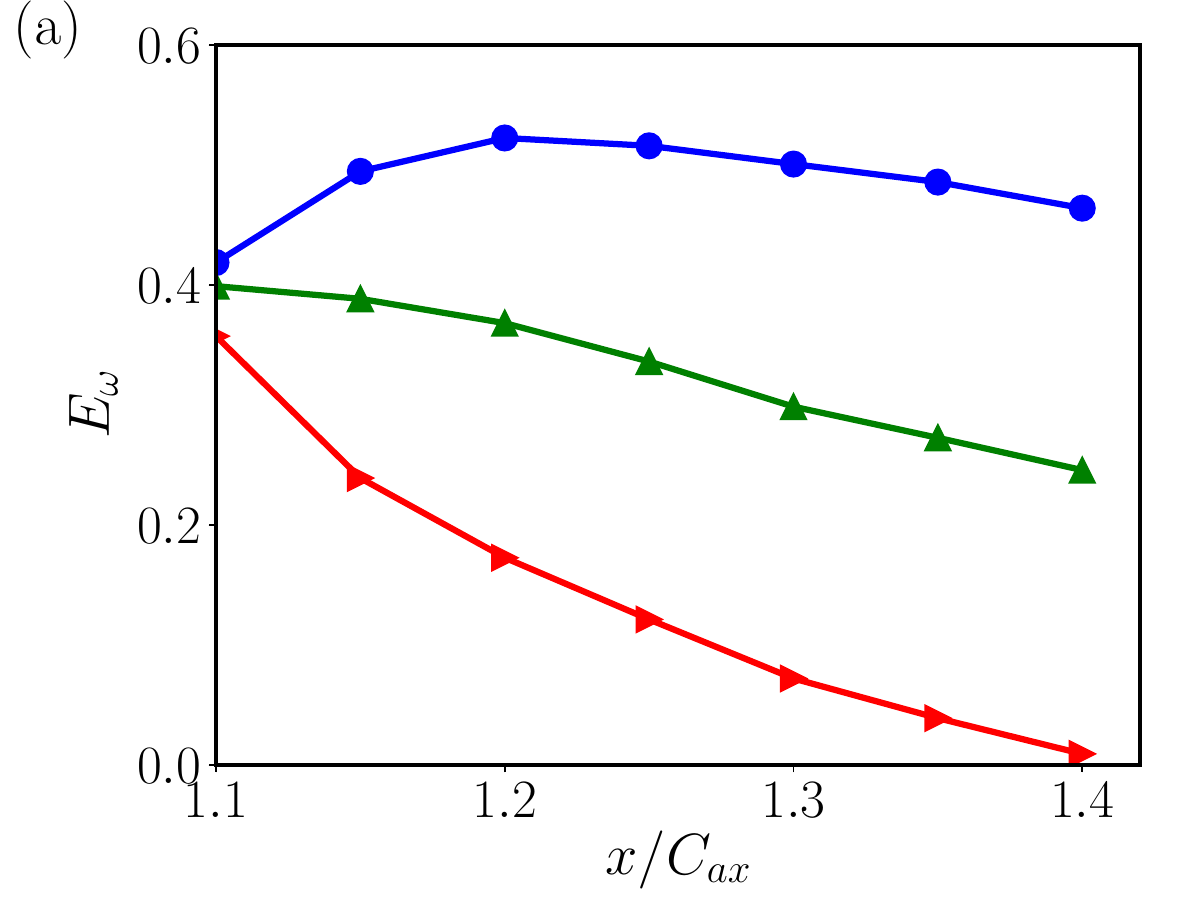}
    \end{minipage}}
  \centering \subfigure{
    \begin{minipage}[c]{0.48\textwidth}{}
    \includegraphics[width=2.8in]{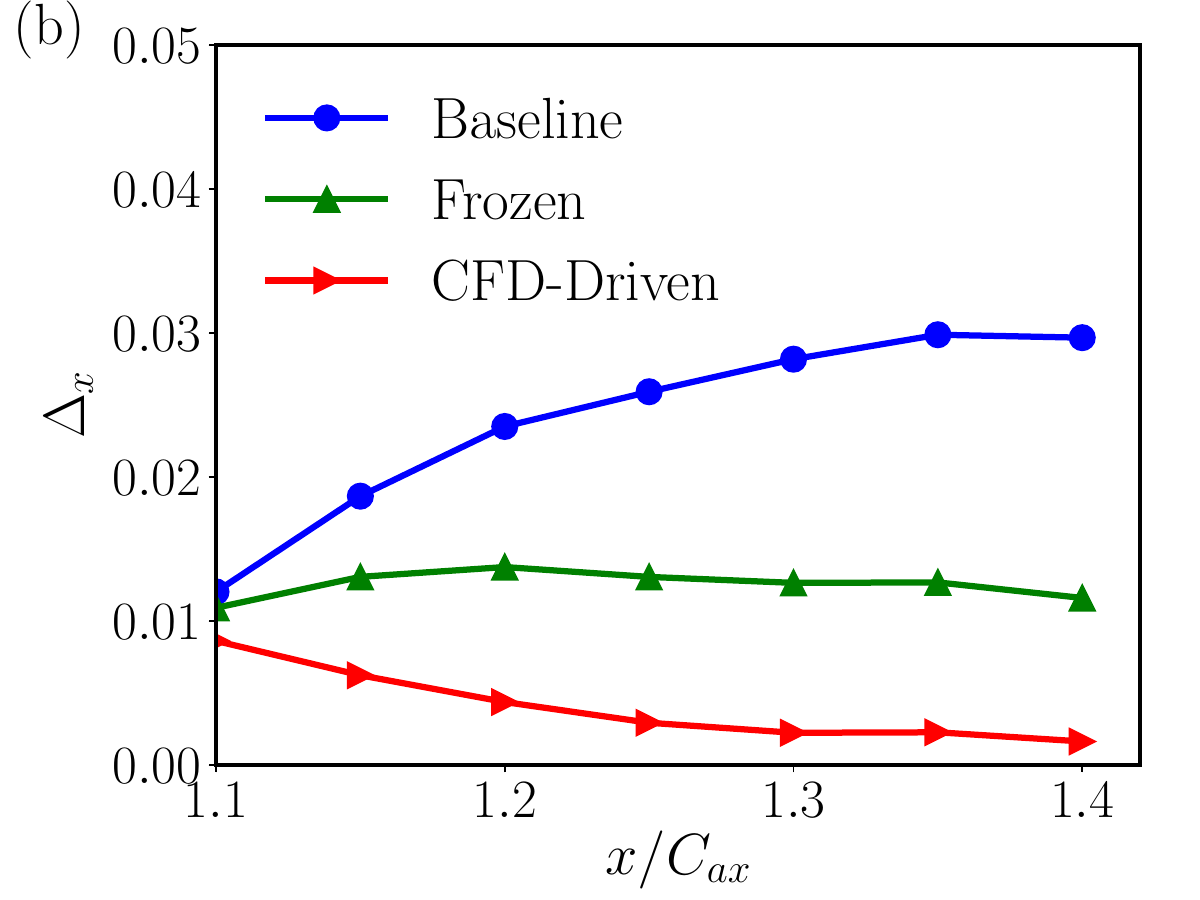}
    \end{minipage}}
    \caption{Characterization of the wake profile evolution at different streamwise locations downstream the blade trailing edge: (a) the peak loss error $E_{\omega}$; (b) the normalised deviation between the wake loss profiles from Hi-Fi and RANS results $\Delta_x$.}
  \label{fig:HPT_wake_dev}
  \end{figure}

To further investigate the development of the wake mixing, the errors of the peak wake loss
\begin{equation}
  E_\omega = \frac{|\omega_{max}^{Hi-Fi}-\omega_{max}^{RANS}|}{\omega_{max}^{Hi-Fi}}
  \label{eq:peak}
\end{equation}
from the different models are ploted against the streamwise coordinate $x$ in Fig.~\ref{fig:HPT_wake_dev}(a).
The CFD-driven model, compared to the baseline and frozen models, siginificantly improves the predictions of peak wake loss at different downstream locations.
Only in the region very close to the traling edge does the error not improve significantly. This is related to vortex shedding exhibiting high-levels of coherence in this region, in which the deterministic unsteadiness inherently violates the RANS assumption and was shown to require a \rthree{different} type of model development \cite{lav2019framework}.
In addition, the evolution of the normalised deviation between the wake loss profiles from Hi-Fi and RANS $\Delta_x$, as defined by Eqn.~(\ref{eq:fitness}), is presented in Fig.~\ref{fig:HPT_wake_dev}(b). The deviations from the Hi-Fi data obtained from the CFD-driven model are remarkably smaller than those obtained from the baseline and frozen training models, showing that the trained model can provide a better prediction for the mixing in the entire wake region.

It is noted that improving predictions for mean flow quantities like kinetic loss in the wake is of critical interest for turbine designers, which is inherently guaranteed by the CFD-driven training as the fittest models are the ones with the smallest cost values.
Improvement in Reynolds stress prediction, however, is also important for capturing the correct physics. As shown in Fig.~\ref{fig:HPT_wake}(b), the CFD-driven model enhances the prediction of the shear stress profile, especially in terms of the suction-side peak value.
Nevertheless, discrepancies exist on the pressure-side, which is presumably because steady RANS has limitations in modelling the coherent vortex shedding under the strong pressure-gradient near the blade trailing edge~\cite[]{Wheeler2016}.

\begin{figure}
  \begin{center}
    \includegraphics[width=0.9\textwidth]{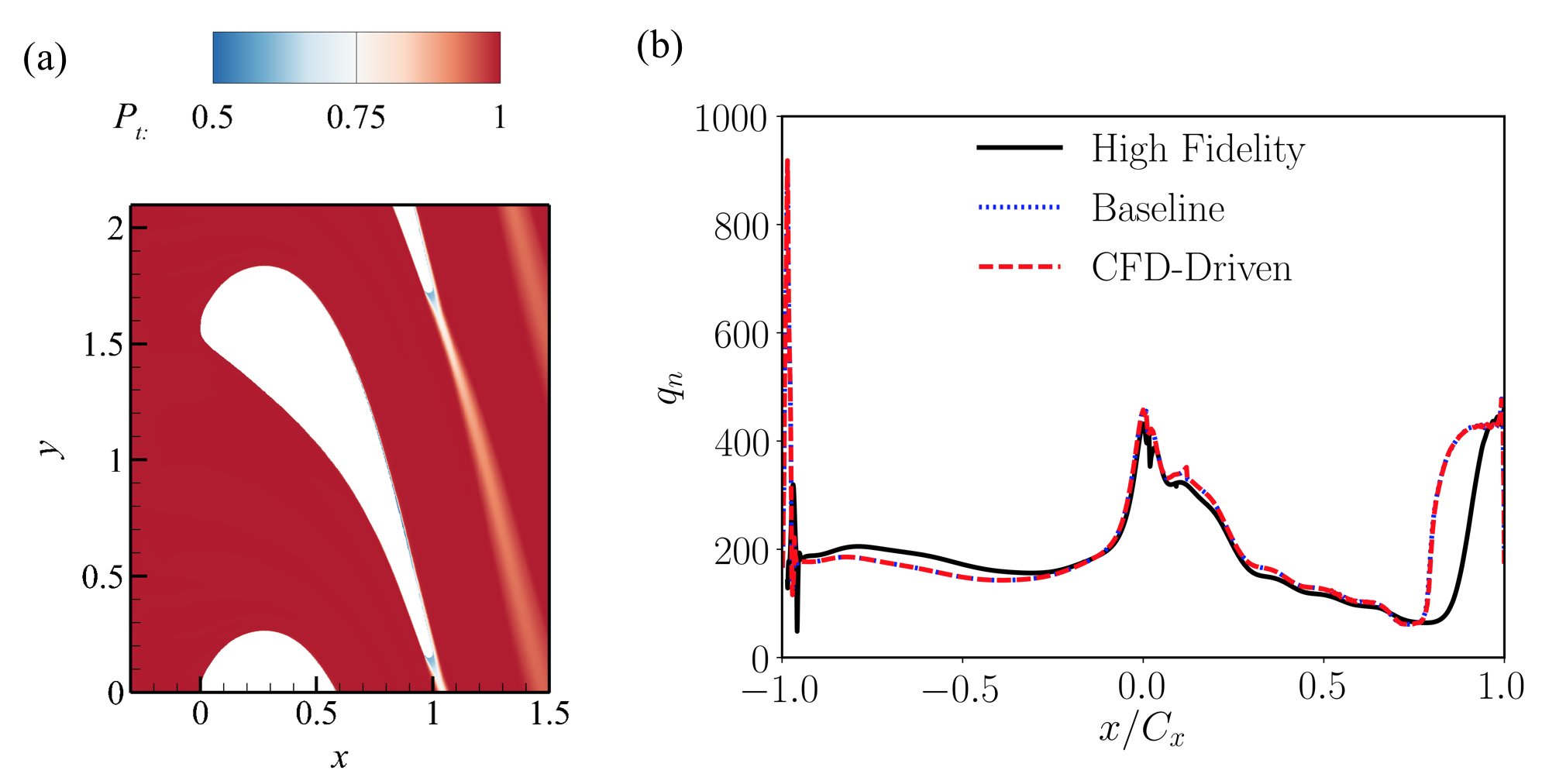}
    \caption{\rone{(a)The contour of the total pressure $P_t$ with the CFD-driven model applied in case A. (b) Wall heat flux around the blade.}}
  \label{fig:GEP_applied}
  \end{center}
\end{figure}

\rone{Moreover, an overview of the flow field obtained by applying the CFD-driven model is shown by the contour of the total pressure in Fig.~\ref{fig:GEP_applied}(a). 
While the trained model is only applied in the masked region as indicated in Fig.~\ref{fig:cases}(b), the switch from the trained model to the baseline model at the boundaries of the wake region works smoothly.
In addition, the wall heat flux around the blade from RANS calculations with both the baseline model and the CFD-trained model are plotted in Fig.~\ref{fig:GEP_applied}(b), showing that applying the trained model in the wake region does not affect the boundary layer flow around the blade. 
It is also noted that the RANS predictions of the blade heat flux deviate from the high-fidelity results in the aft section of the  suction-side boundary layer, which is because the transition is not very accurately captured by the RANS models.}

\subsection{Realizability of the trained model}
One common concern with model development by machine learning might be the realizability of the trained model.
Nevertheless, as the polynomials are based on the invariant tensor basis $V_{ij}^k$, the models trained in the present study are all Galilean invariant~\cite{pope1975more}, which can be considered as one advantage of the symbolic regression via GEP.
To further investigate the realizability of the trained models, tensor analysis for the Reynolds stress anisotropy, like Lumley triangle~\cite{lumley1977return}, is usually applied.
In the present study, the barycentric map developed by Banerjee \emph{et al}.~\cite{banerjee2007presentation} is used to examine the trained model, to provide a nondistorted visual representation of the turbulence anisotropy~\cite{wang2017physics}.

\begin{figure}
  \begin{center}
    \includegraphics[width=0.6\textwidth]{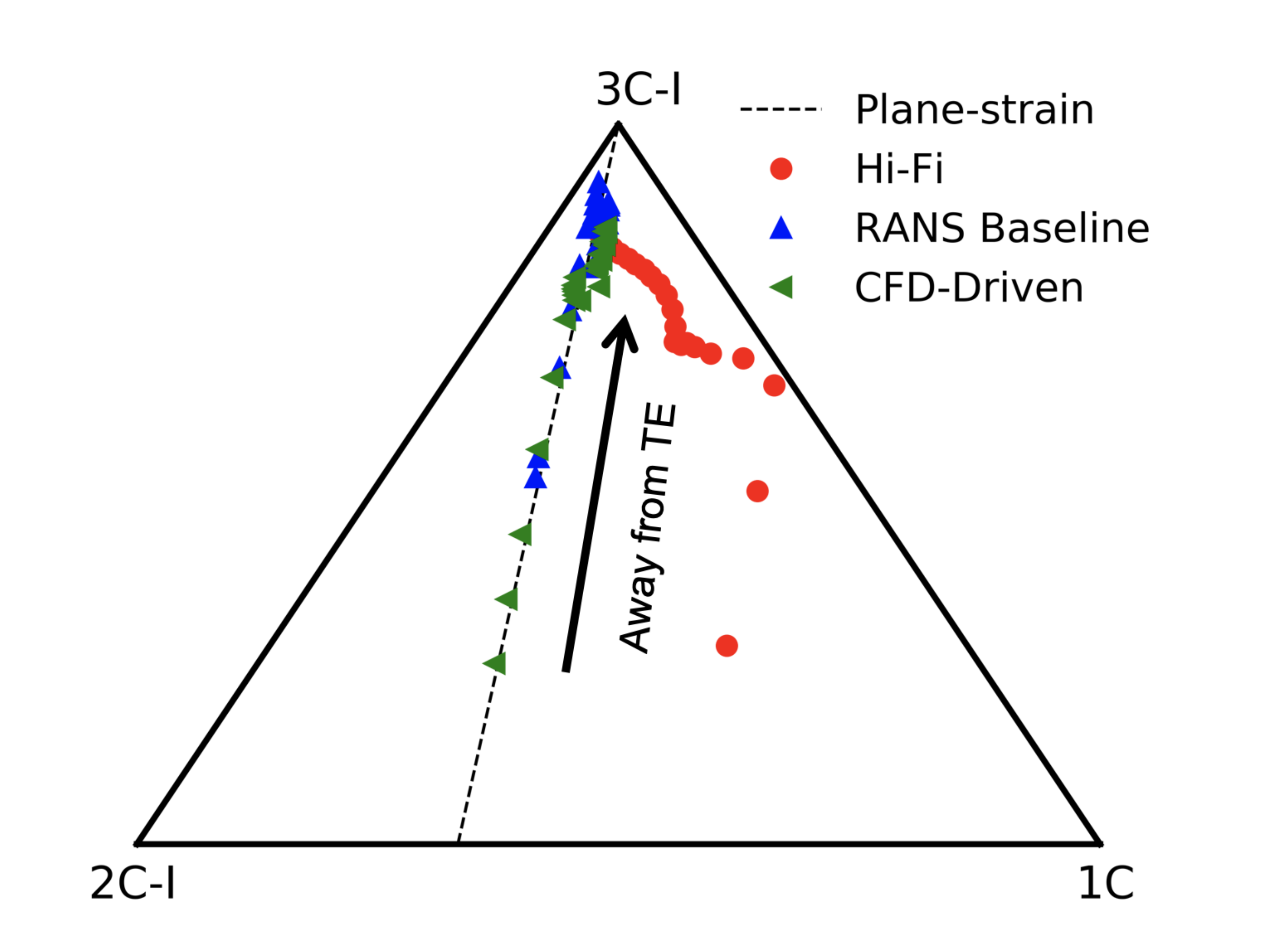}
    \caption{Barycentric map for the Reynolds stress anisotropy in the wake mixing region, from the high-fidelity simulations (Hi-Fi), the $k-\omega$ SST as baseline model (RANS baseline), and the model trained using the CFD-driven method (CFD-driven).}
  \label{fig:bary_map}
  \end{center}
\end{figure}

To get the barycentric map, the eigenvalues of the anisotropy tensor $a_{ij}$ are first obtained as $\left\{\lambda_1,\lambda_2,\lambda_3\right\}$, with $\lambda_1\geq\lambda_2\geq\lambda_3$. Then the barycentric coordinates $(C_{1c},C_{2c},C_{3c})$ can be calculated as
\begin{equation}
  \begin{aligned}
  C_{1c} &= \lambda_1 - \lambda_2,\\
  C_{2c} &= 2(\lambda_2 - \lambda_3),\\
  C_{3c} &= 3\lambda_3 + 1.
  \label{eq:bary_coords}
  \end{aligned}
\end{equation}
With $(\xi_{1c},\eta_{1c})$, $(\xi_{2c},\eta_{2c})$, and $(\xi_{3c},\eta_{3c})$ representing the coordinates of the vertices of the triangle as shown in figure~\ref{fig:bary_map}, the coordinates of the data point $(\xi,\eta)$ in the barycentric map can be given as
\begin{equation}
  \begin{aligned}
  \xi &= C_{1c}\xi_{1c}+C_{2c}\xi_{2c}+C_{3c}\xi_{3c},\\
  \eta &= C_{1c}\eta_{1c}+C_{2c}\eta_{2c}+C_{3c}\eta_{3c}.\\
  \label{eq:xi_eta}
  \end{aligned}
\end{equation}
It is noted that the three vertices in the map represent special states of the anisotropy tensor, including the three-components isotropic (3C-I), the two-components isotropic (2C-I), and the one-component (1C) state.

For the CFD-driven model applied in the HPT case A, the Reynolds stress anisotropy is extracted in the wake mixing region as shown in figure~\ref{fig:HPT_wake}. The barycentric map of the CFD-driven model, along with the baseline $k-\omega$ SST model and the Hi-Fi data, is presented in figure~\ref{fig:bary_map} for comparison.
Moving away from the blade trailing edge, the Reynolds stress anisotropy in the wake region approaches the three-components limit for all cases. This is because the turbulence gradually develops to be isotropic after the strong anisotropic state near the trailing edge.
It is also noted that the RANS stresses show significant difference compared to the Hi-Fi results.
The reason is that while the Hi-Fi simulation is capable of resolving the complex flow physics including the vortex shedding, the steady RANS calculations fail to model such low-frequency deterministic unsteadiness. The RANS flow fields in the wake are dominated by the strong shear layer, and thus the Reynolds stress anisotrpy in the barycentric map is close to the plane-strain limit with $\lambda_2=0$~\cite{banerjee2007presentation}, which is represented by the black dashed line in figure~\ref{fig:bary_map}.

Moreover, due to the extra terms in Eqn.~(\ref{eq:model_HPT}), the CFD-driven model introduces stronger anisotropy and thus has larger-amplitude eigenvalues, which means
\begin{equation}
  \begin{aligned}
  \lambda_1^{CFD} &> \lambda_1^{Baseline},\\
  \lambda_3^{CFD} &< \lambda_3^{Baseline}.
  \label{eq:lambda}
  \end{aligned}
\end{equation}
With $\lambda_2\approx0$ for both RANS calculations, it is straightforward to deduce that the \rthree{coordinates} for the CFD-driven model are relatively further from the three-components isotropic limit in the barycentric map, compared to the same points from the baseline model.

\section{Discussion}
\subsection{Cross-validation of trained model}
\begin{figure}
  \centering \subfigure{
  \begin{minipage}[c]{0.3\textwidth}{}
  \includegraphics[width=2.0in]{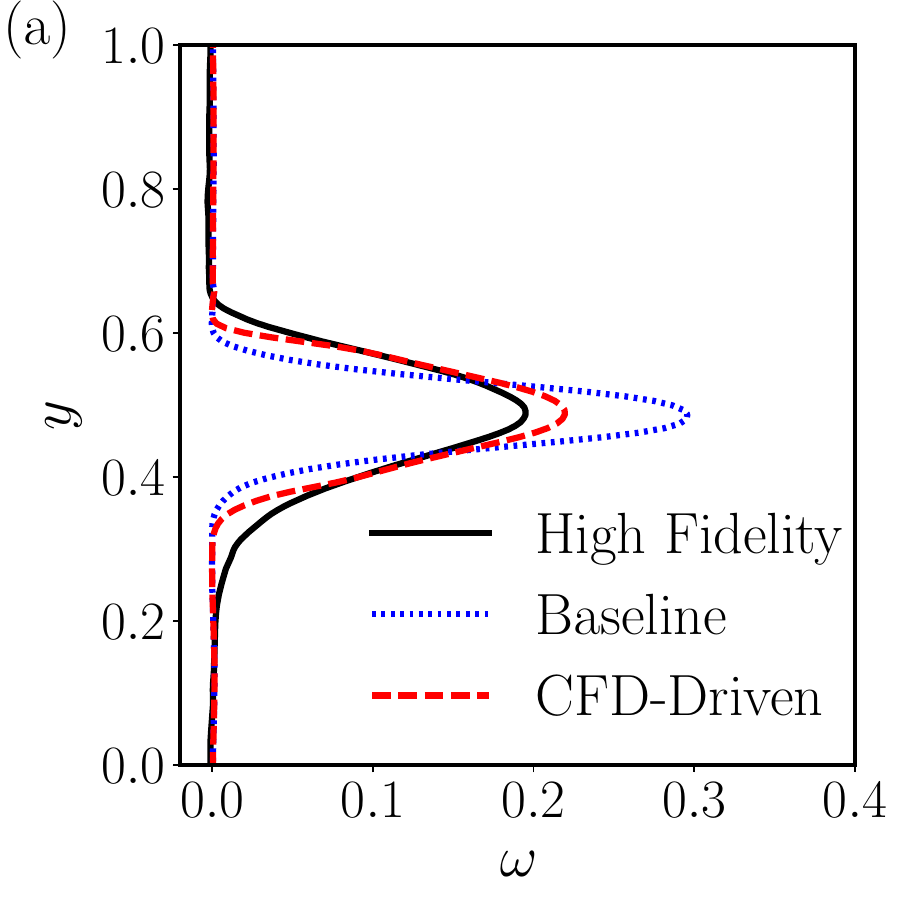}
  \end{minipage}}
\centering \subfigure{
  \begin{minipage}[c]{0.3\textwidth}{}
  \includegraphics[width=2.0in]{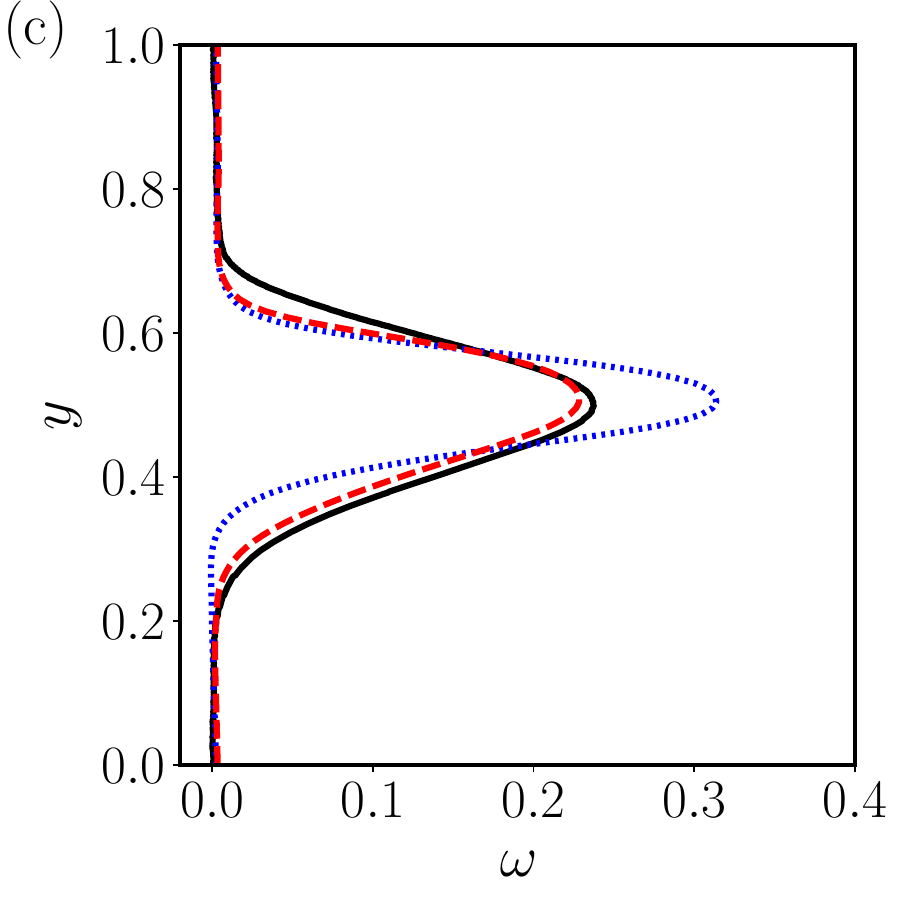}
  \end{minipage}}
\centering \subfigure{
  \begin{minipage}[c]{0.3\textwidth}{}
  \includegraphics[width=2.0in]{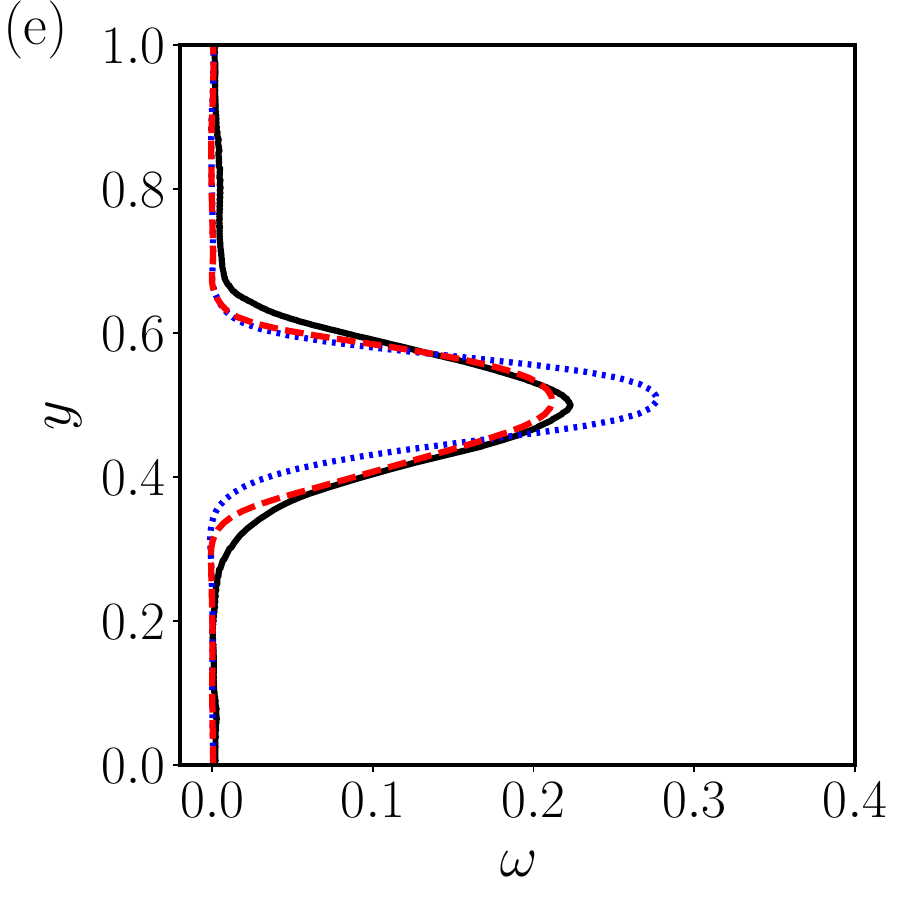}
  \end{minipage}}
\centering \subfigure{
  \begin{minipage}[c]{0.3\textwidth}{}
  \includegraphics[width=2.0in]{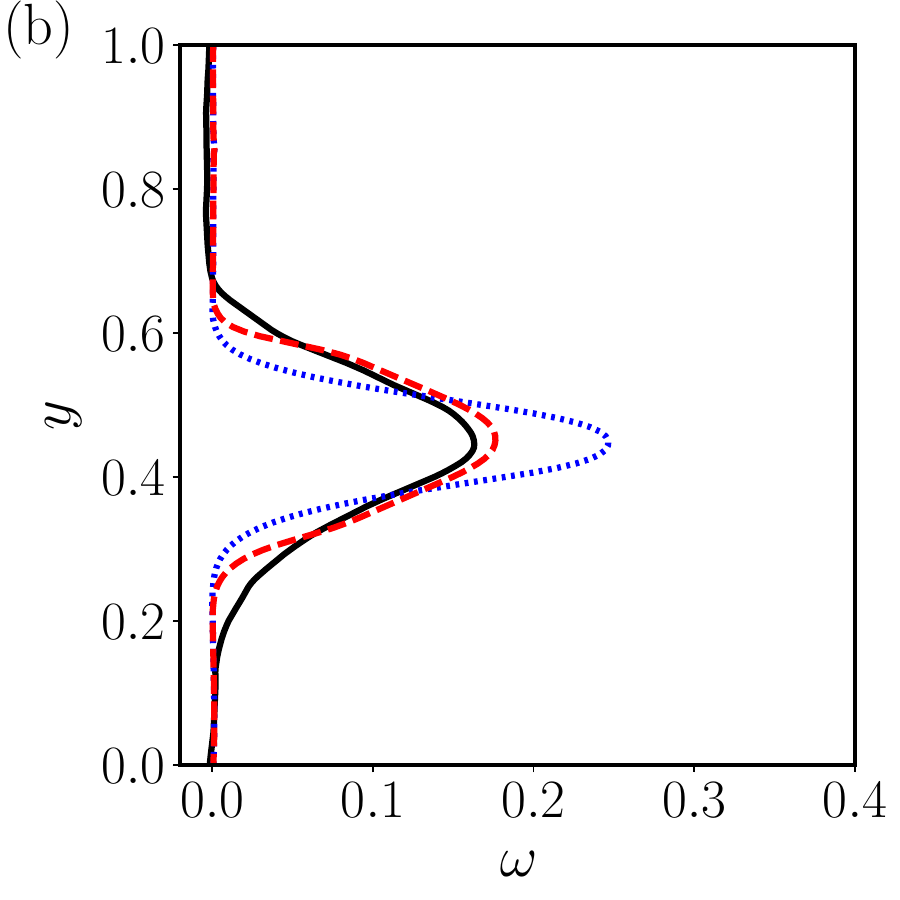}
  \end{minipage}}
\centering \subfigure{
  \begin{minipage}[c]{0.3\textwidth}{}
  \includegraphics[width=2.0in]{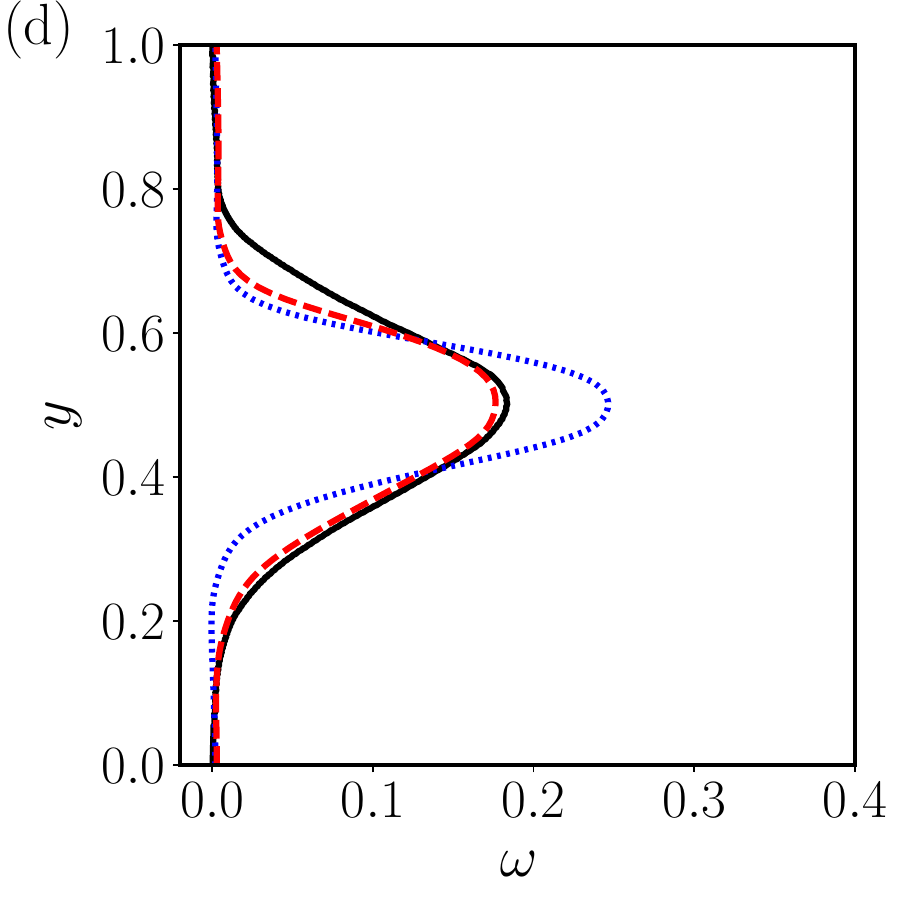}
  \end{minipage}}
\centering \subfigure{
  \begin{minipage}[c]{0.3\textwidth}{}
  \includegraphics[width=2.0in]{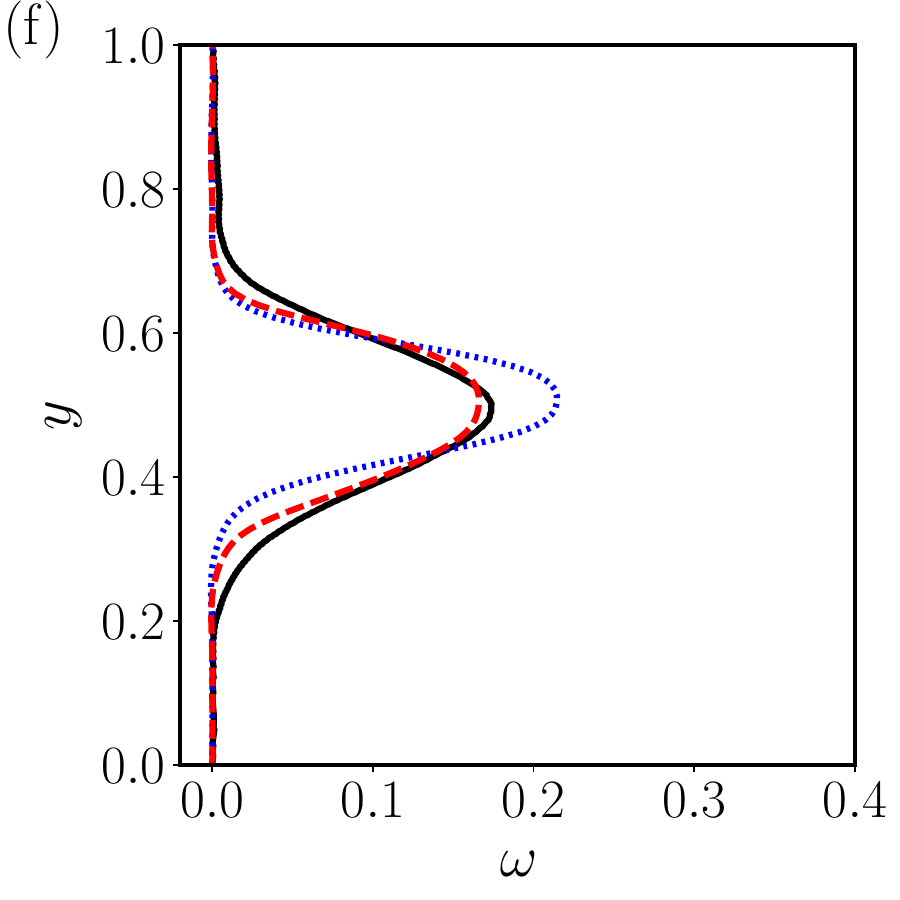}
  \end{minipage}}
  \caption{Kinetic loss profiles in the wake from test cases in table \ref{tab:cases}, (a) HPT case B, $20\%C_{ax}$ downstream; (b) HPT case B, $30\%C_{ax}$ downstream; (c) LPT case C, $20\%C$ downstream; (d) LPT case C, $30\%C$ downstream; (e) LPT case D, $20\%C$ downstream; (f) LPT case D, $30\%C$ downstream.}
  \label{fig:testing}
\end{figure}

It has been presented in Sec.~\ref{sec:result} that the model trained with the CFD-driven method shows much improved predictive accuracy in wake mixing in the \textit{a posteriori} test, and the analysis also demonstrates that the model is physically realizable.
The next step, therefore, is to investigate whether similar accuracy improvements can be achieved for cases with a broad parameter space.
In the present section, the CFD-driven model in Eqn.~(\ref{eq:model_HPT}) is now applied to the three test cases listed in table \ref{tab:cases}, including the HPT case B featuring a higher Reynolds number, and the LPT cases with very different geometries, inflow angles, Reynolds numbers, Mach numbers, and thus flow physics.

The kinetic loss profiles from different downstream locations from all of the three cases are presented in Fig.~\ref{fig:testing}, with the Hi-Fi data and baseline RANS available for comparison.
As expected, the baseline RANS calculations based on the Boussinesq hypothesis result in over-prediction of the wake peak and under-prediction of the wake width across the different cases.
In contrast, the CFD-driven model presents encouragingly accurate prediction of the wake profiles in all test cases due to the additional diffusion and anisotropy added in the Reynolds stress.
Considering the test cases cover a reasonably wide parameter space, this cross-validation underscores the applicability and generality of the CFD-driven trained model in wake mixing flows with diverse flow parameters, geometries and key flow features.

\subsection{Generalizability of the CFD-driven training}
\label{sec:discussion}
As discussed in Sec.~\ref{sec:model}, the trained model focuses on the wake mixing problems so that it is only applied in the masked wake region, while the baseline $k-\omega$ SST model is applied in the rest of the flow domain.
This is called zonal treatment which has a long history in RANS modelling~\cite{matai2019zonal}.
The idea is that we should not expect a model with a unified set of parameters to show good performance for different kinds of flow phenomena, including laminar boundary layer, separation, transition, wake mixing, etc.
A more realizable way is to tune model parameters for specific flow regions, which has already been widely \rthree{accepted} by industrial designers.

While we do not expect the model in the present study to show universally good performance for the whole domain, 
the trained model has been extensively validated in wake mixing cases at different Reynolds and Mach numbers, which are representative of the modern turbine nozzles in industrial applications.
Therefore, the trained model promises to be applicable in wake regions in turbomachinery flows.

Although the model presented here is limited to a specific class of flows, the CFD-driven training framework shows its potential to be applied for general model development efforts.
As discussed in Duraisamy \emph{et al}.~\cite{duraisamy2018turbulence}, one major obstacle for model developement in complex flows is the lack of appropriate training data, especially the high-order statistics like Reynolds stresses which are usually unavailable from experiments.
Compared to the traditional training which mainly focuses on fitting the Reynolds stresses from Hi-Fi data, the CFD-driven training evaluates and optimises models based on integrated CFD calculations.
By comparing the RANS results to Hi-Fi data, the cost function in CFD-driven training is flexible as it can be defined based on any important flow feature of interest to designers.
In particular, mean flow quantities at limited location of the flow domain, like the wake profiles in the present study, can now be used to train models.
Therefore, the CFD-driven training can be applied in more complex flows due to the flexibility of defining the cost function.

The computational cost introduced by the integrated CFD calculations in the training process, however, so far limits the applicability of CFD-driven training.
Typically, the candidate models in GEP need to iterate through hundreds or thousands of generations to get the desired models, and for each generation the CFD claculations have to be performed. Therefore, it is essential that the integrated CFD solver is highly efficient, so that the computational cost is acceptable.

\section{Conclusions}
\label{sec:conclusion}
A novel machine learning framework named CFD-driven training, based on the GEP method, is introduced to develop turbulence models.
As the candidate EASM-like models are explicitly given via symbolic regression in GEP, the model equations can be implemented into a RANS solver, and the cost functions of the models are evaluated by running CFD in the training loop.
By integrating the RANS calculations, the CFD-driven training is able to develop models adapted to the RANS environment and straightforward to be implemented into existing RANS solvers.

The CFD-driven framework was applied to one HPT case to train an EASM-like model for wake mixing in turbomachines.
Compared to the traditional model training, the CFD-driven training produces a model that is in a relatively simple form without numerically stiff terms, and thus shows reliable applicability to practical RANS calculations.
Furthermore, the generated model is tested \textit{a posteriori} for the training case and three other cases covering different flow configurations.
Due to the extra diffusion introduced by the CFD-driven model, the results show overall very good agreement with the kinetic loss profiles obtained from Hi-Fi data in all cases.

While the model trained in the present study is limited to wake mixing, the CFD-driven machine learning is shown to be a promising framework for general turbulence model development. Nevertheless, introducing RANS calculations in the model development iterations increases the computational cost of training over the frozen approach. 
\rone{The relatively high computational cost might limit the applicability of the CFD-driven machine learning to more complex cases such as fully three-dimensional flows. 
However, this could be offset by efforts in reducing the computational cost in future studies, and possible directions include analyzing the population size, optimizing the convergence of the evolutionary algorithm, and improving the efficiency of the CFD solver through introducing advanced technologies like GPU acceleration.}

\section*{Acknowledgments}
	This research used resources of the Oak Ridge Leadership Computing Facility, which is a DOE Office of Science User Facility supported under Contract DE-AC05-00OR22725.
	This work was supported by a grant from the Swiss National Supercomputing Centre (CSCS) under project ID s884.
	This work was also supported by the resources provided by the Pawsey Supercomputing Centre with funding from the Australian Government and the Government of Western Australia.

\section*{Appendix A. Grid convergence study}
\label{sec:appedix_A}
\begin{figure}
  \centering \subfigure{
    \begin{minipage}[c]{0.48\textwidth}{}
    \includegraphics[width=2.8in]{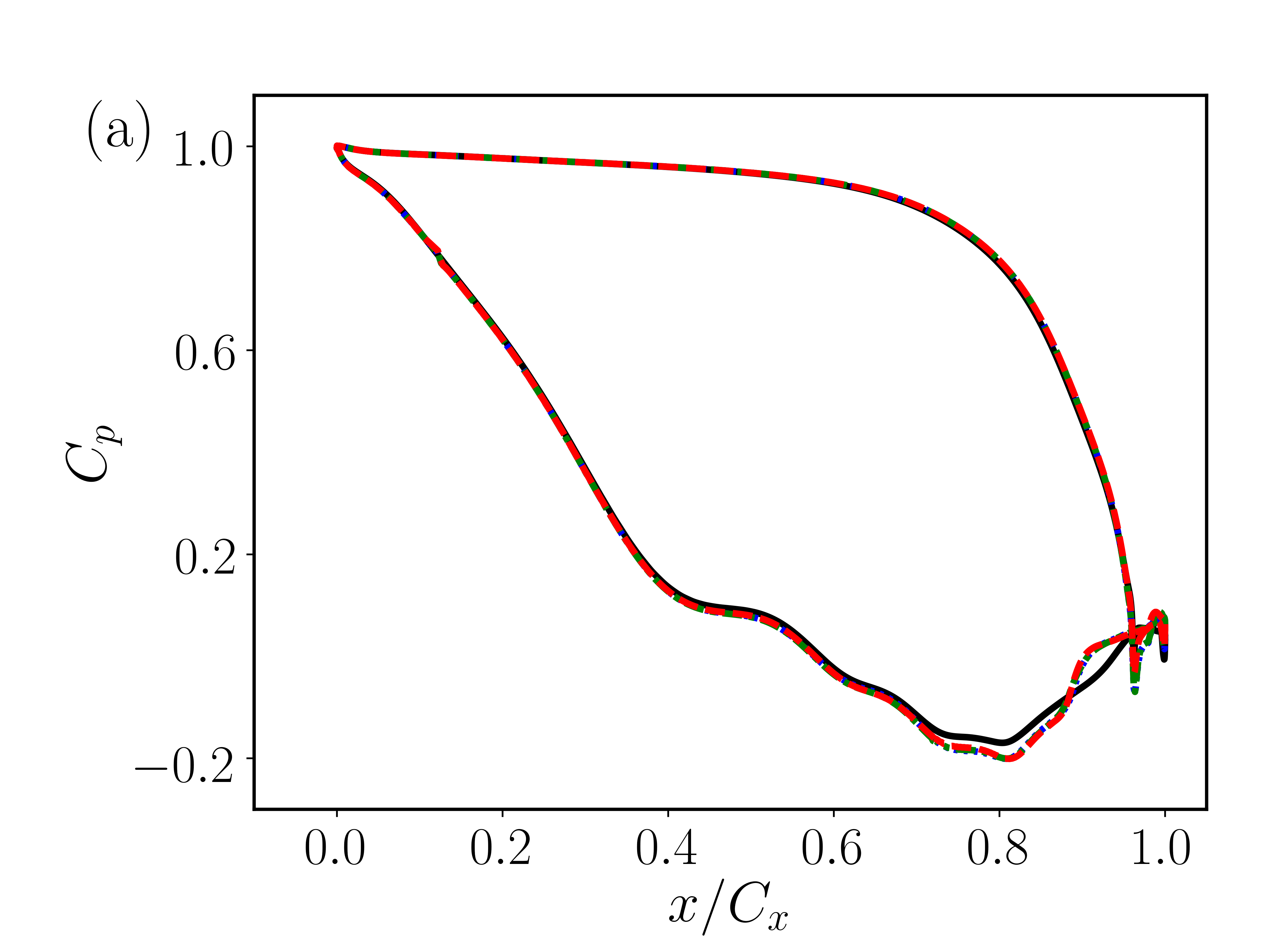}
    \end{minipage}}
  \centering \subfigure{
    \begin{minipage}[c]{0.48\textwidth}{}
    \includegraphics[width=2.8in]{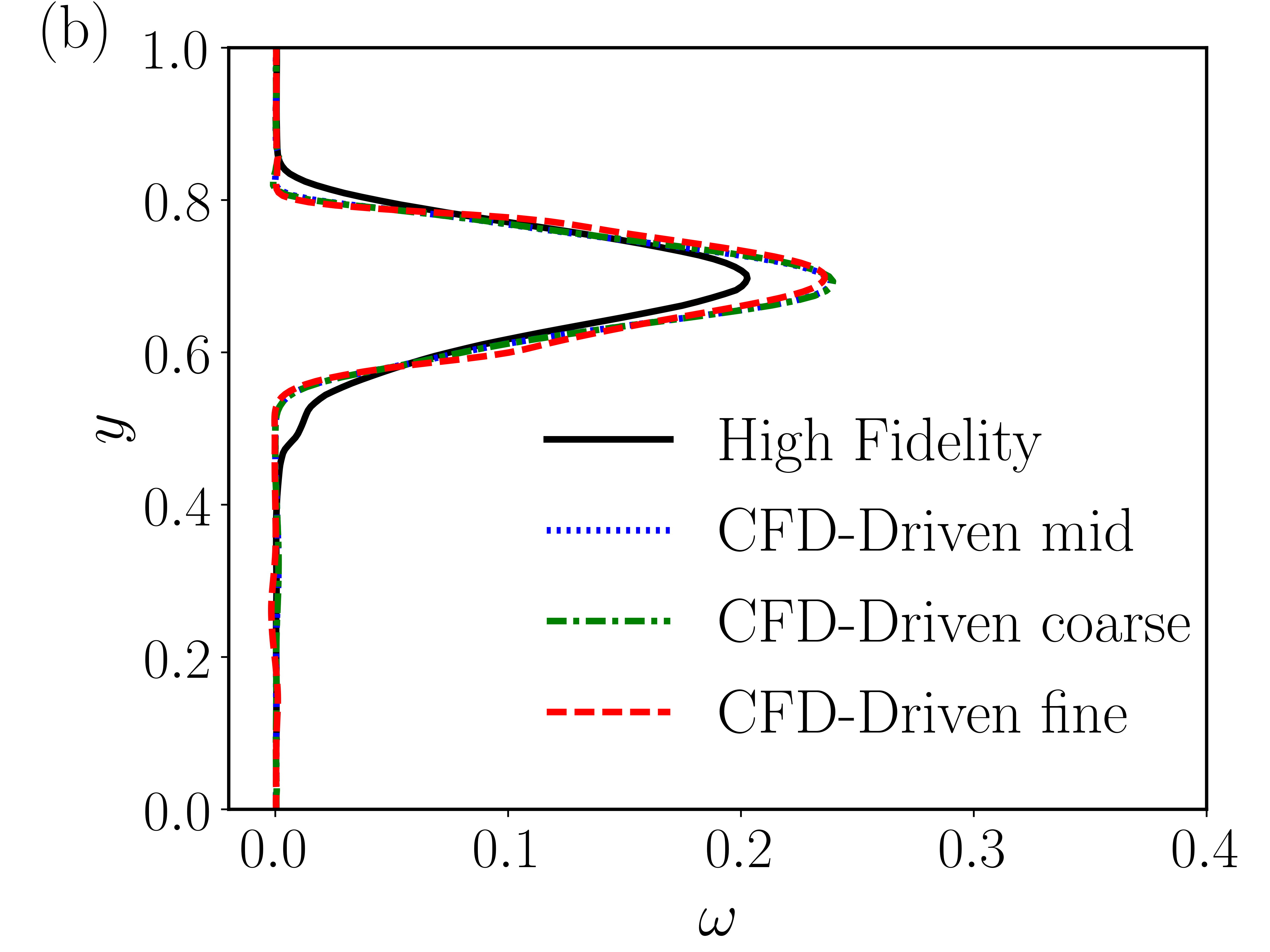}
    \end{minipage}}
    \caption{Grid convergence study of the HPT case A using the trained RANS model: (a) pressure coefficient distribution around the blade; (b) kinetic wake loss profile at $20\%C_{ax}$ downstream of the trailing edge.}
  \label{fig:grid_convergence_inloop}
\end{figure}

\begin{table}[t]
\caption{RANS grid convergence parameters for case A.}
\begin{center}
\label{tab:grid}
\begin{tabular}{c l l l}
& & \\ 
\hline
Grids & O-Grid & H-Grid & Total Points \\
\hline
Coarse & $565\times57$ &$81\times101$ & 40,386 \\
Mid & $705\times65$ & $101\times161$ & 62,086 \\
Fine & $1081\times113$ & $161\times320$ & 173,673 \\
\hline
\end{tabular}
\end{center}
\end{table}
\rone{
The mesh used for the model development in the present study was selected on the basis of a grid-refinement analysis. The dimensions for the three different grids to establish grid independence are listed in Tab.~\ref{tab:grid}.
The pressure coefficients $C_p$ obtained from RANS calculations applying the CFD-driven trained model are plotted versus fraction of axial chord and compared against the LES data in Fig.~\ref{fig:grid_convergence_inloop}(a). The results show good agreement between the RANS calculations with different grid resolutions.
Furthermore, Fig.~\ref{fig:grid_convergence_inloop}(b) shows the kinetic wake loss profiles at $20\%C_{ax}$ downstream of the blade trailing edge.
We can see that both the pressure distribution around the blade and the kinetic wake profiles from the RANS calculations are grid-independent.
}

\section*{Appendix B. Sensitivity of the wake region selection}
\begin{figure}
  \centering \subfigure{
    \begin{minipage}[c]{0.48\textwidth}{}
    \includegraphics[width=2.8in]{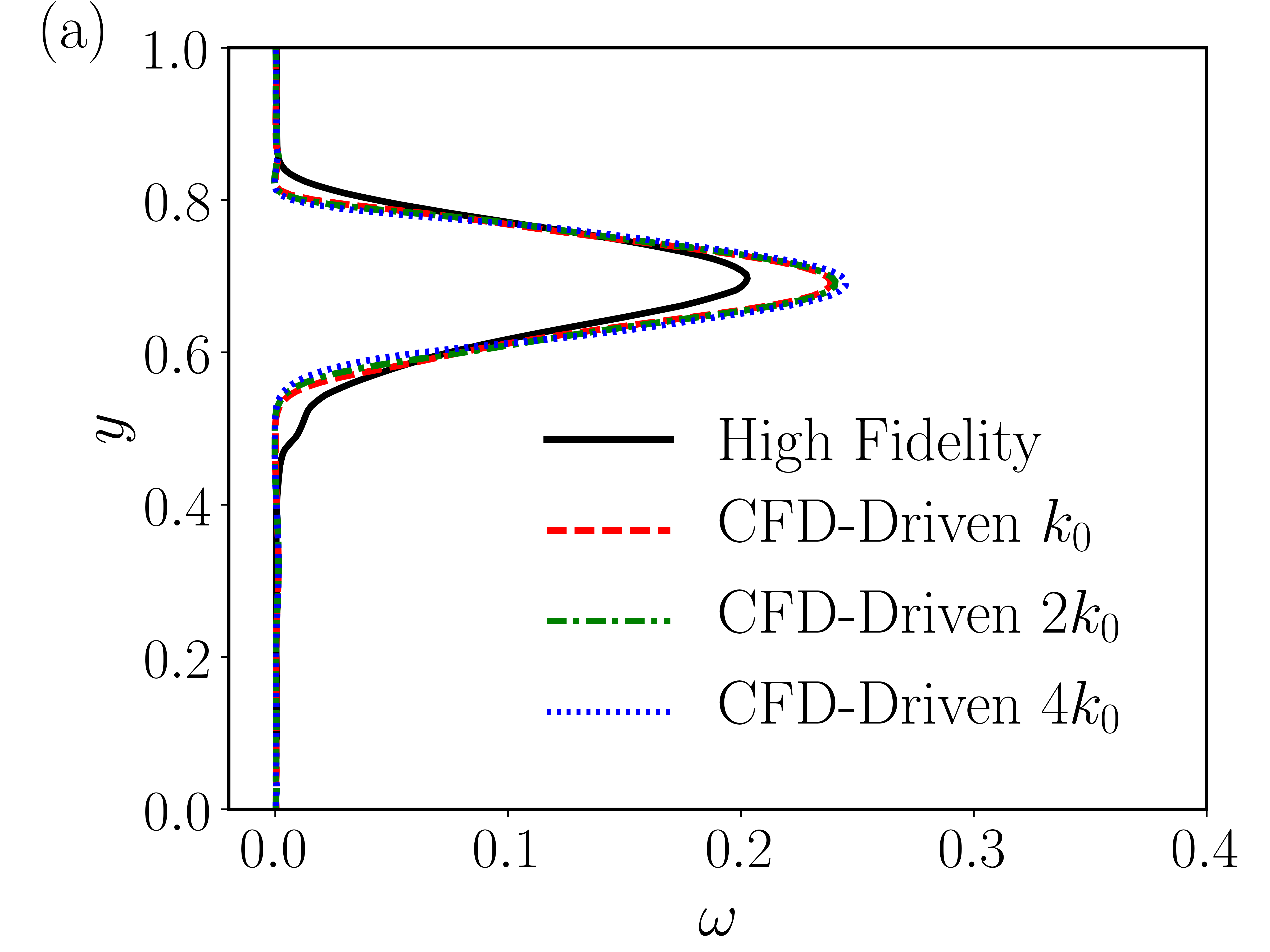}
    \end{minipage}}
  \centering \subfigure{
    \begin{minipage}[c]{0.48\textwidth}{}
    \includegraphics[width=2.8in]{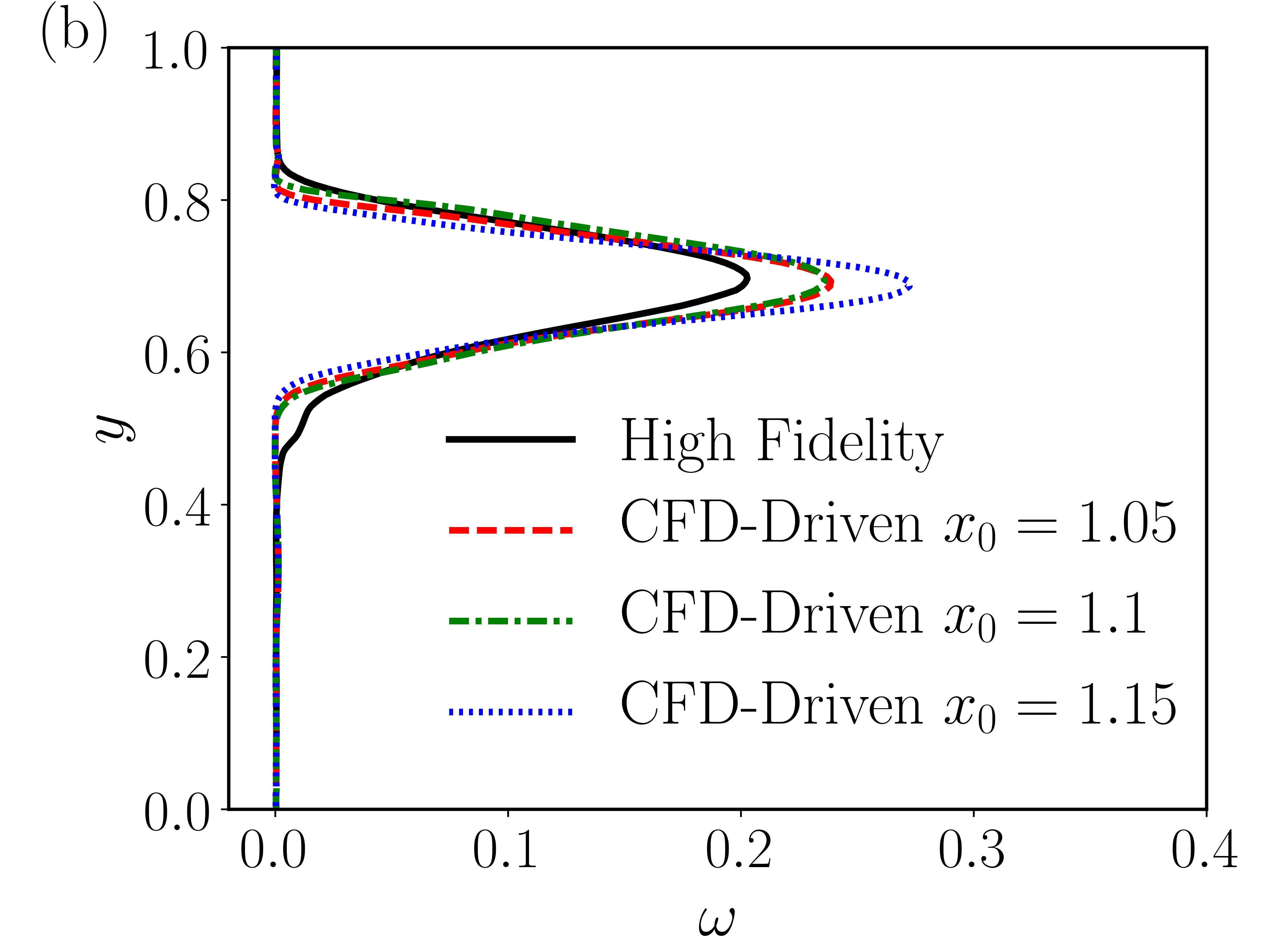}
    \end{minipage}}
    \caption{Kinetic wake loss profiles at $20\%C_{ax}$ downstream of the trailing edge: (a) the CFD-driven model applied in wake region masked by different levels of $k$; (b) the CFD-driven model applied in wake region masked by different $x$ locations.}
  \label{fig:window_sensitivity}
\end{figure}

\rone{
It is noted that the trained models are only applied in the wake region, which is extracted based on the criteria $k>5\%k_{max}$ and $x>1.05$ as shown in Fig.~\ref{fig:cases}(b). To further investigate the sensitivity of the selection of this wake region, we have varied the level of TKE threshold and the $x$ location, and the results are presented in Fig.~\ref{fig:window_sensitivity}.
It is shown that changing the TKE level from $k_0=5\%k_{max}$ to $4k_0$ does not significantly affect the wake profiles, which is because the TKE is concentrated in the wake region and varying the threshold does not significantly change the extracted region.
Furthermore, moving the wake region downstream from $x>1.05$ to $x>1.1$ does not significantly affect the wake profile at $x=1.2$, while a training region with $x>1.15$ is simply not large enough so that the wake prediction is affected by upstream effects. 
}

\bibliographystyle{model1-num-names}
\bibliography{cfd_in_loop}

\end{document}